\documentclass[runningheads]{llncs}

\usepackage{graphicx,paralist,wrapfig,xspace,amsmath,comment,color}

\usepackage[colorinlistoftodos]{todonotes}

\usepackage{booktabs}
\usepackage{longtable}
\usepackage{booktabs} % For formal tables
\usepackage{algorithm}
\usepackage{amsfonts}
\usepackage[noend]{algpseudocode}
\usepackage{flushend}
\usepackage{multirow}
\usepackage{listings}
\usepackage{xcolor}
\usepackage{enumitem}
\usepackage{comment}
\usepackage{footmisc}
\usepackage{longtable}
\usepackage{wrapfig}
\usepackage{lipsum}
\setitemize{noitemsep,topsep=0pt,parsep=0pt,partopsep=0pt}
\usepackage{array,booktabs,arydshln,xcolor}
\usepackage{tikz}
\usetikzlibrary{tikzmark}
\newcommand{\satire}{{\sffamily\scshape Satire}\xspace}

\begin{document}
%
%\title{Symbolic Abstraction-guided Technique for
%       Rigorous Floating-point Error analysis}
%\title{\large Scalable Abstraction-guided Technique for Incremental Rigorous analysis
%of round-off Errors}
%
%\title{\large Scalable Abstraction-guided Technique for \\ Rigorous
%              Floating-Point Error Analysis}

\title{\large An Abstraction-guided Approach to\\
Scalable and Rigorous Floating-Point Error Analysis }

\titlerunning{\satire: Abstraction-guided Rigorous Floating-Point Error Analysis}

%\author{First Author\inst{1}\orcidID{0000-1111-2222-3333} \and
%Second Author\inst{2,3}\orcidID{1111-2222-3333-4444} \and
%Third Author\inst{3}\orcidID{2222--3333-4444-5555}}

\author{Arnab Das\inst{1} \and
Ian Briggs\inst{1} \and
Ganesh Gopalakrishnan\inst{1} \and
Pavel Panchekha\inst{1} \and
Sriram Krishnamoorthy\inst{2}}


\institute{University of Utah, USA \and
Pacific Northwest National Laboratory, USA}
%\author{}
%\institute{}

%\author{Arnab Das\inst{UoU}, Sriram Krishnamoorthy}
%\institute{%
%  {University of Utah, PNNL}
%}

%\author{Sriram Krishnamoorthy}
%\institute{%
%  {Pacific Northwest National Laboratory}
% }

\maketitle              % typeset the header of the contribution
\begin{abstract}
  
Automated techniques for rigorous floating-point round-off error analysis
are important in areas including
formal verification of correctness and precision tuning.
Existing tools and techniques, while providing tight bounds,
fail to analyze expressions with more than a few hundred
operators, thus unable to cover important practical problems.
In this work, we present \satire, a new tool that
sheds light on how scalability and bound-tightness
can be attained through
a combination of incremental analysis, abstraction, and
judicious use of concrete and symbolic evaluation.
\satire has handled problems exceeding 200K operators.
We present \satire's underlying error analysis approach,
information-theoretic abstraction heuristics,
and a wide range of case studies, with evaluation
covering FFT, Lorenz system of equations, and various PDE stencil types.
Our results demonstrate the tightness of  \satire's bounds, its
acceptable runtime, and valuable insights provided.

  \keywords{Floating-point arithmetic; Rigorous Round-off analysis;
    Symbolic Differentiation; Abstraction; Scalable Analysis}
\end{abstract}

%-----------------------------------------------------------------
\section{Introduction}
\label{sec:intro}
Floating-point arithmetic underlies many critical
scientific and engineering applications.
Round-off errors introduced during floating-point
computations have resulted in egregious errors
in applications ranging from statistical
computing and econometric software~\cite{economic,numstats} to critical
defense systems~\cite{patriot-failure}.
To guard against such errors, formal verification tools are essential, and
a basic capability in such tools is the ability to
{\em rigorously} estimate the round-off error generated by
floating-point expressions for given input intervals.
This capability is also important in designing
rigorous mixed-precision tuning systems
(e.g., \cite{fptuner}).

A number of tools for rigorous round-off analysis have been
developed: Fluctuat~\cite{fluctuat}, Gappa~\cite{gappa},
PRECiSA~\cite{precisa}, Real2Float~\cite{MagronCD15},
Rosa~\cite{rosa}, FPTaylor~\cite{DBLP:journals/toplas/SolovyevBBJRG19},
Numerors~\cite{DBLP:conf/sas/JacqueminPV18}, 
and tools specialized for cyberphysical systems
\cite{munoz-winding,rocco-solver-nfm-2019} to name a few.
Unfortunately, these tools
cannot {\em automatically} handle floating-point
expressions of thousands of operators
that arise in even the simplest of practical situations
including recurrence equation solving schemes and
algorithms such as the Fast Fourier Transform (FFT) and %, and other familiar
%algorithms such as
parallel prefix sum.
Today's tools in this space
can at best handle expressions with a few dozen operators.
To tackle large sizes automatically, one must today employ
abstract interpretation methods and/or
theorem-proving
methods~\cite{fluctuat,munoz-winding,munoz-gappa,boldo-2nd-order-wave}.
The use of non-rigorous
approaches that employ concrete evaluation on a reasonable sampling
of the input space~\cite{sc13-precimonious,hifptuner,herbie,craft,adapt}
or Monte-Carlo-style approached~\cite{verificarlo,stoke-fp} is
also popular.
While such approaches are ultimately inevitable,
formal verification is essential for
critical tasks such as 
aircraft collision avoidance verification~\cite{munoz-gappa-pvs}.
A higher degree of automation can greatly help with
these verification tasks.

An analysis method's ability to derive tight bounds is as important as
its ability to handle large expressions graphs.
Interval analysis provides an efficient and fast way to 
obtain error bounds; however, it inherently loses correlation
between terms, resulting in quite pessimistic bounds.
While affine analysis maintains correlations
between error variables by keeping them formal (symbolic), 
it can also result in  pessimistic bounds, especially for
non-linear expressions such as $x/(x+1)$, where the non-linearity
is introduced by division.

In this work, we introduce a scalable and rigorous approach
for error analysis embedded in a new tool, namely,
\satire({\bf S}calable
{\bf A}bstraction-guided {\bf T}echnique for {\bf I}ncremental {\bf R}igorous
analysis of round-off {\bf E}rrors). \satire supports scalable mixed-precision analysis and multi-output
estimation of straight-line
floating-point code.
We demonstrate that \satire can perform error estimation for
complex expression graphs involving thousands of operators, including
FFTs, recurrence equations, and partial differential equation (PDE) solving.

%while interval analysis~\cite{interval-anal}
%is very scalable, it can give quite pessimistic bounds
%(e.g., $(x-x)$ can return $[-2,2]$ when given $x\in [-1,1]$).
%%
%While affine analysis~\cite{affine-anal} maintains correlations
%between variables (and returns $0$ as the result of $(x-x)$),
%it also provides pessimistic bounds, especially for
%non-linear expressions such as $x/(x+1)$ where the non-linearity
%is introduced by division.

The recently proposed method of rigorously computing
{\em symbolic Taylor forms} and embodied in the
FPTaylor tool~\cite{DBLP:journals/toplas/SolovyevBBJRG19}
avoids problems associated with interval
and affine methods.
Moreover, as the table
of results in~\cite{DBLP:journals/toplas/SolovyevBBJRG19}
and \cite{DBLP:conf/sas/JacqueminPV18} show,
FPTaylor also produces some of the tightest bounds when compared
to the tools in its class.
However, FPTaylor cannot handle large expression graphs.
For example, for the discretized one-dimensional heat flow
(``1D-heat'') equation
calculation using a stencil whose temporal application
is unrolled over 10 time steps, about 442 expression nodes are
generated.
FPTaylor takes about 8 hours simply to generate the Taylor form expression,
following which its global optimizer fails to handle this expression.
In contrast, \satire handles the aforesaid 1D-heat problem
in a few seconds, generating tight bounds.
\satire also has handled expressions with over 200K operators
drawn from realistic examples.
In addition, \satire\ generates tighter error bounds than
FPTaylor on most of FPTaylor's examples.
To give an early indication of \satire's performance,
Table~\ref{tab:small-comparison} provides a sampling of examples
(comprehensive evaluations in \S\ref{sec:contributions}
and \S\ref{sec:eval}).

%% -- adding a small comparative table early on --
\begin{table}[htbp]
\centering
\begin{tabular}{cccccccc}
\toprule
\multirow{2}{*}{Benchmarks} & \multirow{2}{*}{\# operators} & \multicolumn{2}{c}{Exec Time (secs)} & \multicolumn{2}{c}{Absolute Error} &
\raisebox{-4ex}{\shortstack[l]{Empirical Max\\ Abs Error}}\\
\cmidrule(lr){3-4} \cmidrule(lr){5-6}
& & {\satire} & {FPTaylor} & {\satire} & {FPTaylor} \\
\midrule
jetEngine 			& 35 	& 1.969   & {\bf 0.84}	& {\bf 2.88E-08}	& 7.09E-08 & 1.57E-12\\
kepler2	  			& 43 	& {\bf 0.402}   & 0.52	&  3.20E-12		& {\bf 2.53E-12} & 3.49E-13\\
dqmom	  			& 35 	& {\bf 0.206}   & 0.63	& {\bf 9.66E-10} & 6.91E-05  & 3.27E-13 \\
1D-heat-10steps		& 442 	& {\bf 2.1} & - & 6.11E-15 & NA & 5.45E-16 \\
FFT-1024pt 			& 43K & {\bf 180} & - & 7.80E-13 & NA & 6.18E-15\\
Lorenz-20steps		& 307 	& {\bf 284} & - & 1.25E-14 & NA & 3.59E-15 \\
%fdtd1d-64steps		& 192k & {\bf 9711} & - & 7.45e-13 & NA & 9.89e-16\\
\end{tabular}
\label{tab:small-comparison}
\end{table}

%In this work, we provide a scalable
%and rigorous approach for error analysis embedded in a new tool
%\satire ({\bf S}calable
%{\bf A}bstraction-guided {\bf T}echnique for {\bf Incremental} {\bf R}igorous
%analysis of round-off {\bf E}rrors)
%that outperforms FPTaylor on virtually all of FPTaylor's benchmarks.
%
%
%--

The benchmarks Jetengine, Kepler, and DQMOM are some of the largest
that have been handled by FPTaylor, but contain no more than a few dozen
operators.
The much larger benchmarks including
FFT-1024pt (an FFT computation of 1K inputs),
and
Lorenz-20steps (a Lorenz system that captures the ``butterfly effect'')
%and fdtd1d-64steps (finite-difference time domain used in electromagnetic
%simulation and hundreds of other applications)
are all representative of practically important examples.
They far exceed the capability of all existing round-off error analysis
tools.
The last column of the table presents
the empirically calculated (using shadow values)
{\em maximum} absolute error over
$10^6$ simulations for randomly selected
data points from the input
intervals.
Lacking other means of easily obtaining tight rigorous bounds,
these results provide a good indication
of how close \satire's bounds
are.

\satire only analyzes for first-order error.
To check whether we are lacking precision due to this,
we ran FPTaylor on its own benchmarks with its second-order
error analysis turned off, only finding that the resulting error
values were still bit-identical.
This shows that it can be very difficult to
create examples where second-order error matters.
In~\cite{aiken-math-dot-h}, the authors further emphasize
this point by mentioning that they had to refine their
second-order error analysis beyond the approach taken in
FPTaylor.

\vspace{-1.6ex}
\paragraph{Roadmap:\/}
In \S\ref{sec:contributions}, we explain \satire's contributions
at a high level; this helps better understand the remainder of the paper.
In \S\ref{sec:contributions}, we explain the essential background
on floating-point arithmetic and the
global optimization-based error analysis done in \satire.%
We also point out the key contributions made by \satire.
In \S\ref{sec:illustrative-example}, we explain the
incremental analysis adopted in \satire.
\S\ref{sec:error-analysis} presents  the formal
details of \satire's error analysis with heuristics
for abstraction and for improving global optimization/canonicalization
detailed in \S\ref{sec:heuristics}.
\S\ref{sec:eval} presents a comprehensive evaluation of
\satire on many practical examples.
\S\ref{sec:relwork} presents additional related work.
\S\ref{sec:conc} has our concluding remarks.

%--end

%-----------------------------------------------------------------
\section{Background, Contributions}
\label{sec:contributions}
A binary floating-point number systems, $\mathbb{F}$, is a
subset of real numbers representable
in finite precision and expressed as tuple $(s, m, e, p)$
%--Here, $\beta=2$ corresponds to the radix-2 floating-point system,
with $p=53$ and $24$ for double and single precision numbers,
respectively,
$s \in \{-1,1\}$ is the sign bit, $m$ the 
mantissa, and $e$ the exponent.
We consider only normal numbers where $m\in [1,2)$.
%
%There are two types of floating-point
%numbers, the {\em normals} and {\em denormals}, other than the special cases
%of NaNs and infinities.
%Normal numbers lie in the interval
%$[1, 2)$, while denormals are in $[0, 1)$.
%
%---For double-precision numbers, this range is $[-1022,1023]$. %$-1022 \leq e \leq 1023$.
%
Any such number, $\tilde{x} \in \mathbb{F}$
has the value $s \cdot m \cdot 2^e$.
If $x \in \mathbb{R}$, then $\tilde{x}$ denotes an element in $\mathbb{F}$
closest to $x$ obtained by applying the rounding operator($\circ$) to x.
The IEEE 754 standard defines four rounding modes for elementary floating-point
operations. Every real number $x$ lying in the range of $\mathbb{F}$ can be
approximated by a member $\tilde{x} \in \mathbb{F}$ with a relative error
no larger than the unit {\em round-off} $\bf{u}=\dfrac{1}{2}2^{1-p}$.
Here, $2^{1-p}$ represents the {\em unit of least precision (ulp)} for 
the exponent value of 1. Thus, to relate the quantities $\tilde{x}$ and $x$
over the rounding operator,
	if $x \in \mathbb{R}$ lies in the range of $\mathbb{F}$, then
$\tilde{x} = \circ(x) = x(1 + \delta),\quad |\delta| \leq \boldsymbol{u}.$

%Based on the radix and precision, the round-off quantity varies.
%We support mixed-precision analysis, by suffixing the rounding
%operator, when necessary, to denote the precision for the rounding.
%%
%For example, when rounding
%in double-precision we denote $\circ_{64}$.
%
Given two exactly represented floating-point numbers, $\tilde{x}$ and $\tilde{y}$,
arithmetic operators $\diamond \in \{+,-,\cdot,/\}$ have the following guarantees:
\vspace{-0.5ex}
\begingroup
\small
\begin{equation}
\label{eq:opbound}
\begin{split}
	\circ(\tilde{x} \diamond \tilde{y}) &= (\tilde{x} \diamond \tilde{y})(1 + \delta),\quad |\delta| \leq \boldsymbol{u} \\
	&= (\tilde{x} \diamond \tilde{y}) + (\tilde{x} \diamond \tilde{y})\delta 
	= \{ \hbox{real value} \} + \{ \hbox{ round-off error } \}
\end{split}
\end{equation}
\endgroup

Thus, the rounded result involves an interval centered around the real exact value. The
width of this interval is determined by the amount of round-off error accumulated.
Besides the elementary operators, \satire supports other complex mathematical
operations such as transcendentals. In such cases, the bound on the $\delta$ terms changes
as a multiple of $\bf{u}$. We keep a default configurable value for each operator,
allowing the user to obtain customized error bounds
for the specific math library being used.

Affine arithmetic~\cite{affine} is
often employed to keep errors correlated.
A real number $\hat{x}$ is represented in affine form as
$\hat{x} = x_0 + \sum_{i=1}^n x_i \epsilon_i$,
where $x_0$ denotes the central value
and the $x_i$'s are finite floating-point numbers representing
noise coefficients
associated with the corresponding noise variables,
$\epsilon_i$'s. The $\epsilon_i$ are
formal variables whose value lies in $[-1,1]$
but are unknown until assigned.
This representation
allows converging paths to cancel out error terms and improve
tightness.
For example $(x-x)$ in
affine analysis
will yield 0 because the noise variables can cancel
each other out.
Unlike affine analysis, interval analysis keeps widening the intervals.
\satire's approach can be viewed as ``symbolic affine''
(much like FPTaylor) in that
the noise coefficients,
$x_i$, are {\em symbolic}, as
further clarified in \S\ref{sec:symb-affine}.

%--

We use the following notations and conventions in the upcoming sections.
We assume that
an expression ${\cal E}$ defined over inputs ${\bf X}$
is being analyzed for the round-off error, 
given that each $x_i \in {\bf X}$
is instantiated with input interval $I_i \in {\bf I}$.
Here, ${\cal E}$ is assumed to be represented as a DAG, and
the process of analyzing ${\cal E}$ consists of somehow obtaining
the error at each interior DAG node $i$, represented by ${\cal E}_i$,
and then calculating how much this error contributes to the final error.
Suppose ${\cal E}{rr}_i$ be the error generated
by subexpression ${\cal E}_i$ (details in \S\ref{sec:error-analysis}).
%
% If we sum all the ${\cal E}{rr}_i$s, then given that
% these are symbolic expressions, we will obtain the
% same type of error cancellations as provided by standard
% affine arithmetic, except that these cancellations are now
% defined across the input domain.
%
We also use the notation
$E_i$ and $E$ (respectively) to denote
${\cal E}_i$ and ${\cal E}$ that have been evaluated to
ground values (fully evaluated after variable substitutions)
by instantiating $x_i \in {\bf X}$ with intervals $I_i \in {\bf I}$.
We now explain
how \satire ensures scalability (Page~\pageref{sec:symb-affine})
and tight bounds (Page~\pageref{sec:tight-bounds}).
%
%Impact on practice is highlighted in \S\ref{sec:eval}.

\label{sec:symb-affine}
\paragraph{\bf Contributions Toward Scalability: Incremental Approach\/}
\satire adopts an {\em incremental} approach
while generating the error contributions of node ${\cal E}_i$ to
the final output.
In other words, we perform a breadth-first
walk of the expression DAG of ${\cal E}$,
enumerating subexpression nodes ${\cal E}_i$.
At each ${\cal E}_i$, we 
perform first-order error analysis (see \S\ref{sec:error-analysis})
of the total error generated at ${\cal E}_i$ to 
compute the error expression  ${\cal E}{rr}_i$.
We then multiply this error with the {\em path
  strength} of {\bf all the paths} going
  from ${\cal E}_i$ to ${\cal E}$.\footnote{The
  notion of ``path-strength''
is explained in \S\ref{sec:error-analysis}, but basically
it is the value of the derivative of ${\cal E}$ with respect to
${\cal E}_i$ that is calculated using symbolic reverse-mode
automatic differentiation.}
Let there be $M$ such paths with path strengths $p_1, p_2,\ldots p_M$.
In many examples such as 1D-heat,
$M$ grows exponentially 
with the depth of ${\cal E}_i$ due to path reconvergence
in expression DAGs.

In FPTaylor, the Symbolic Taylor Form ends
  up having the form $\Sigma_{j=1}^{M}\; (p_j\cdot {\cal E}{rr}_i)$
  which is huge for large $M$, as the error
  expression ${\cal E}{rr}_i$ is replicated $M$ times.
  This replication happens multiple times, explaining
  why it took 8 hours to generate
  its Symbolic Taylor Form for 1D-heat.

With \satire,
we first compute ${\cal E}{rr}_i$ 
as explained in \S\ref{sec:error-analysis}.
We then compute ${\cal E}{rr}_i \cdot \Sigma_{j=1}^{M}\; p_j$,
which is the product of the error with the {\em effective
path strength} $\Sigma_{j=1}^{M}\; p_j$, taking all the paths
from ${\cal E}_i$ to $E$ into account.
This is significantly smaller in that it does not replicate
  ${\cal E}{rr}_i$ a very large number of times.
  This explains why \satire finished on 1D-heat in a few seconds {\em even
  without using abstractions}.
  Generating such factored forms also aids {\tt SimEngine}~\cite{simengine}
  (detailed below under ``canonicalizations'').

\noindent{\bf Automated Introduction of  Abstractions:\/} During a bottom-up (forward) pass to build ${\cal E}$,
  \satire performs  symbolic execution to build the
  symbolic expression for ${\cal E}_i$. 
While doing so,
\satire uses a heuristic to
decide whether or not to
{\em automatically} create an abstraction
for ${\cal E}_i$.

\begin{figure}
	\centering
    \includegraphics[width=1.0\textwidth]{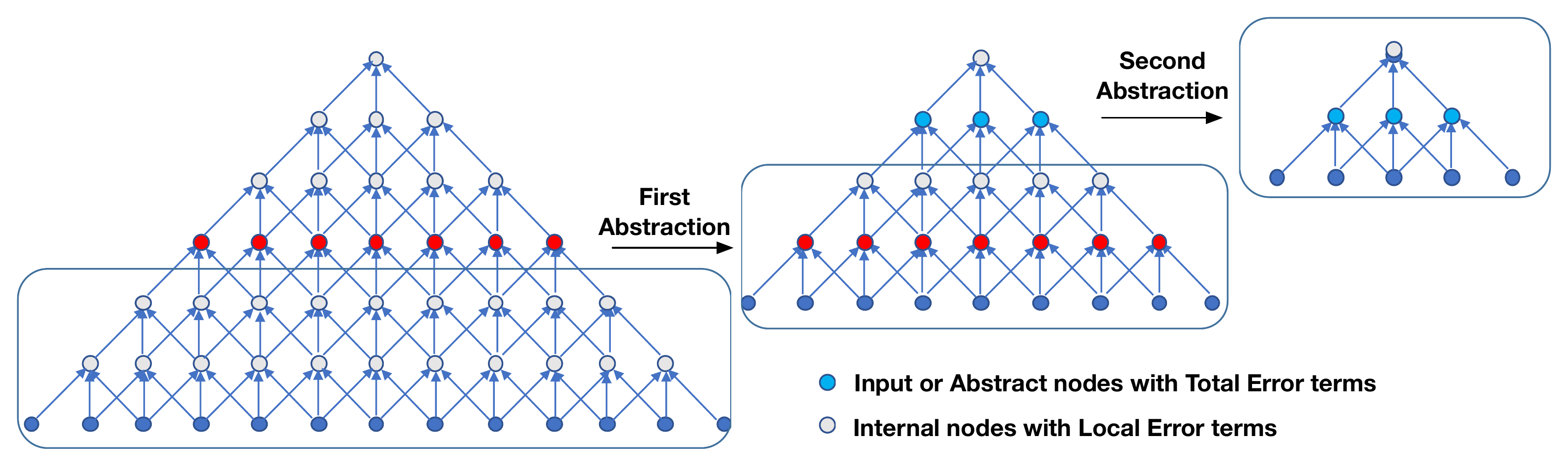}
  \caption{Incremental error analysis using gradual abstraction}
  \label{fig:gradual-abstraction}
\end{figure}
%--

Figure~\ref{fig:gradual-abstraction}
illustrates this incremental abstraction scheme.
We employ a cost metric based on
information-theory~\cite{Shannon1948}
to obtain a set of cut-points
in the abstract syntax tree (AST).
Each of these cut-points serve as the ${\cal E}_i$
for which abstractions are created.
The selected cut-points then become free variables, carrying
concrete error interval bounds (explained below)
for the rest of the error analysis.
In Figure~\ref{fig:gradual-abstraction},
we introduce
our first abstraction at height 2 for a stencil-type
computation.

Clearly there is a risk of losing variable correlations
in the process.
In \S\ref{sec:heuristics},
we point out that with well-chosen heuristics, this risk can
be minimized, providing still tight error estimates.
These abstractions are key to \satire's performance on
larger examples, and can be viewed as an appropriate tradeoff
between scalability and bound tightness.
By performing abstractions,
 the error analysis at each of the
       abstracted nodes ${\cal E}_i$ is a much more manageable process
       and finishes quickly.
       Secondly, we incrementally compute the error bounds of
       ${\cal E}_i$ with respect to the variables ${\bf X}_i$ mentioned
       in it using a global optimization call that refers
       to the appropriate intervals ${\bf I}_i$.
       This is a much smaller optimization query that finishes fast.
       The resulting error expression is a
       concrete value interval that
       is propagated upstream for further error analysis.
       In other words, during \satire's
       analysis, {\em a mixture of
       symbolic and concrete error intervals} are propagated
       across the expression DAG.\footnote{As can be seen,
         \satire is neither ``interval based'' nor
         ``affine based''---it is best viewed as a judicious combination
         of interval and ``symbolic-affine'' analysis.}
       This dramatically reduces overall error analysis complexity.

\label{sec:tight-bounds}
\paragraph{\bf Contributions Toward Tight Bounds:\/}

\noindent{\sl Expression Canonicalization:\/}
A central step that is involved at every stage
in \satire's operation is this: given an
expression ${\cal E}$ and the intervals in which
the variables in ${\cal E}$ lie, compute the
tightest interval bound for ${\cal E}$.
Consider computing this for a simple expression such
as $x\cdot x\cdot x$ using an interval library such
as Gaol where $x\in [-1,5]$.
Gaol would perform two successive interval multiplications,
resulting in $[-25,125]$.
To allow users to obtain better bounds, Gaol provide a
special interface function $pow$ using which the above
call can be written as $pow(x,3)$.
Intuitively the use of functions such as $pow(x,3)$
conveys the information that the same $x$ instance is being
acted upon, allowing Gaol to obtain $[-1,125]$.
Unfortunately, there are only a limited number of
interface functions such as $pow$, and therefore, in general,
an external canonicalizer must be used to reduce the number of
distinct occurrences of variables in an expression.
SimEngine's ``expand'' strategy allows us to achieve this effect,
which then allows the back-end optimizer Gelpia (that uses the
interval branch-and-bound algorithm) to converge to tight bounds much
more quickly.
For the DQMOM example in Table~\ref{tab:small-comparison},
this was the main reason why
we improved upon
FPTaylor's result of {\bf  6.9e-05}
to {\bf 9.6e-10} in \satire (\S\ref{sec:error-analysis}).

\noindent{\sl Avoid Impeding Canonicalization:\/}
Not only must explicit canonicalization be performed
wherever possible, we avoid taking steps that
can block canonicalization.
To be more specific,
denote the sum of path strengths at expression site ${\cal E}_i$,
i.e., $\Sigma_{j=1}^{M}\; p_j$, by ${\cal S}_i$.
When we perform a depth-oriented traversal of the expression DAG,
let us imagine generating expressions
${\cal E}_1, \ldots,  {\cal E}_K$ (where
the ${\cal E}_i$ are the subexpressions of ${\cal E}$)
with corresponding errors
${\cal E}{rr}_1, \ldots,   {\cal E}{rr}_K$.
The final error at the output of ${\cal E}$ is
${\cal E}{rr} = \Pi_{i=1}^{K}\; {\cal S}_i\cdot {\cal E}{rr}_i$.
Let the final {\em concrete} error at the output of ${\cal E}$ be $Err$.
There are two ways to arrive at $Err$: (1)~As a concretization of
  $\Pi_{i=1}^{K}\; abs({\cal S}_i\cdot {\cal E}{rr}_i)$; or (2)~As a concretization of
  $\Pi_{i=1}^{K}\; abs({\cal S}_i) \cdot abs({\cal E}{rr}_i)$.
While both these are real-number equivalent, the latter
expression {\em blocks} canonicalizations from occurring within
{\tt SimEngine}.
All our results jumped from being worse than FPTaylor's
to being better than FPTaylor's when we adopted the former form
(the jump from \satire (UnOpt) to \satire in Table~\ref{tab:satire-fptaylor}).

\begin{comment}
\label{sec:practice}
\paragraph{\bf Contributions Toward Practice:\/}
%
We mention two highlights of our work that directly address practice.
%
The first is
our work on Fast Fourier Transform (FFT) which is very widely used.
%
The best previously reported study on error bounds of FFT that we are
aware of is the conservative analytical bound reported in~\cite{hal-fft}.
%
Using \satire, we are able to produce a much more informative (and tighter)
bound (\S\ref{sec:eval}).
%
Second, scalable error analysis directly helps inform precision-tuning decisions.
%
We report (\S\ref{sec:eval}) a case study on FDTD
(whose stencils have a high condition number~\cite{MullerEtAl2009:HandBook,atomic-condition-su})
to obtain the relative error profile
at runtime using \satire's absolute error bounds.
%
Using a high-precision
shadow value calculation, we show the tighness of the relative error profile,
a direct measure of the number of mantissa bits
lost~\cite{MullerEtAl2009:HandBook}.
\end{comment}

%--end

%-----------------------------------------------------------------
%\section{Background} ABSORBED INTO CONTRIBUTIONS
%\label{sec:background}
%\input{background.tex}

%-----------------------------------------------------------------
\section{Illustrative Example}
\label{sec:illustrative-example}
\begin{figure}[t]
	\centering
    \includegraphics[width=1.0\textwidth]{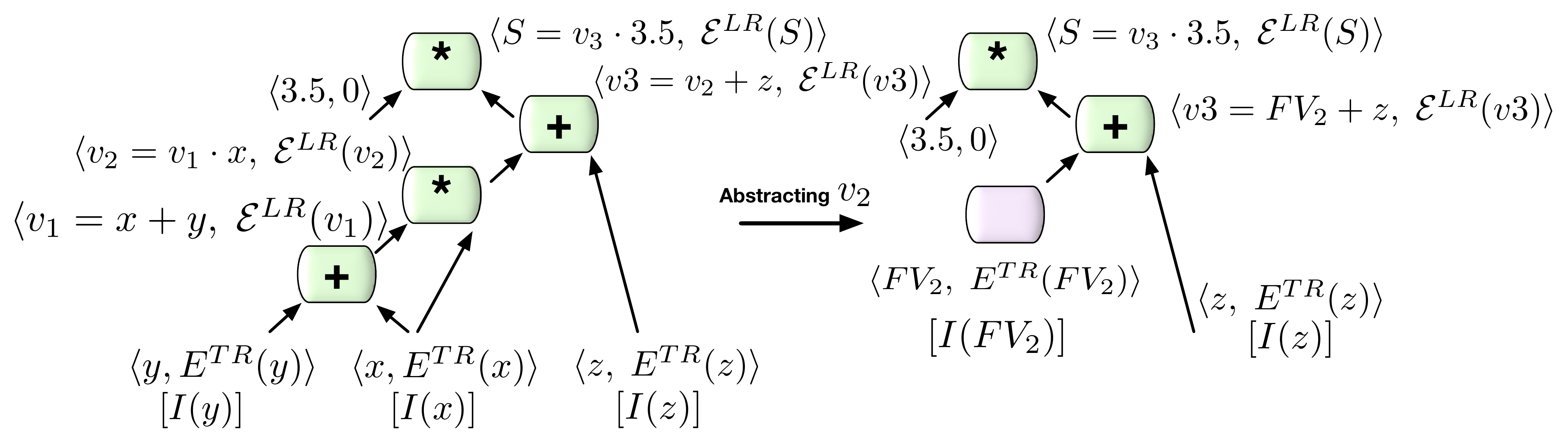}
  \caption{AST for $S = (x\cdot(x+y)+z)\cdot 3.5$, and the rebuilt AST with $v_2$ abstracted.
 The
 equations
 analyzed are $v_1 = x+y;\;\;
 v_2 = v_1\cdot x;\;\; v_3 = v_2+z;\;\; S = v_3 \cdot 3.5$.}
  \label{fig:overview-illustration}
\end{figure}

In \satire, a floating-point straight line program is
parsed into an abstract syntax tree (AST) where each node of the AST
is an operator (e.g., $\{+,-,\cdot,/\}$ or others such as $exp$
 and transcendentals).
%
%--Prior to introducing the
%--formalism, we discuss here with a small example.
Figure~\ref{fig:overview-illustration} shows the AST for the 
expression $S = (x\cdot(x+y)+z)\cdot 3.5$. Here, $x$, $y$, and $z$ are the 
input variables with intervals $I(x)$, $I(y)$, and $I(z)$, respectively.
In ~\cite{rosa,higham}, the authors
introduce the decomposition of errors into {\em generated} round-off
errors and {\em propagation} of incoming errors.
We use  ${\cal E}^{lr}$ to denote the local error term associated with a node.
The total error term for a node, denoted as ${\cal E}^{tr}$, is then the additive 
composition of ${\cal E}^{lr}$ and the propagation of the incoming error,
${\cal E}^{prop}$.\footnote{As before,
 we use ${\cal E}$ for symbolic expressions
 and $E$ for their ground counterparts.}
We annotate
each interior
node in Figure~\ref{fig:overview-illustration} 
with a pair consisting of
its symbolic function expression and
a symbolic local error term.
An abstracted node is treating as an additional input, 
wherein we
take the convention 
of using the variable name as the symbolic
expression for an input node, and instead of the local error, we employ
the {\em total error term} (as detailed below these input
nodes are not modeled).

 %--

Consider node $v_2$ with associated
symbolic expression $v_1 \cdot x$.
Its incoming error is a symbolic expression
denoted by ${\cal E}^{lr}(v_2)$.
From Equation~\ref{eq:opbound}:
\begingroup
\small
\begin{equation}
	\label{eq:internal-node-error}
  \begin{split}
  	 {\cal E}^{tr}(v_2)  = {\cal E}^{lr}(v_2) + {\cal E}^{prop}(v_2) \ 
       & = {\cal E}^{lr}(v_2) +
        \left ( E^{tr}(x)
        \dfrac{\partial v_2}{\partial x}
        + {\cal E}^{tr}(v_1)
        \dfrac{\partial v_2}{\partial v_1} \right )\\
        & =
        {\cal E}^{lr}(v_2)
         + \left ( E^{tr}(x)\cdot v_1 + {\cal E}^{tr}(v_1)\cdot x \right )
  \end{split}
\end{equation}
\endgroup

\noindent The derivatives signify the sensitivity of $v_2$ w.r.t. their operands.
Similarly,
%
%Note that we have a total error term for $v_1$ in 
%
%It can be further unrolled into its local 
%
\begingroup
\small
\begin{equation}
  \label{eq:v1-unroll}
  \begin{split}
   {\cal E}^{tr}(v_1) = {\cal E}^{lr}(v_1) + {\cal E}^{prop}(v_1) 
  & = {\cal E}^{lr}(v_1)
   + \left ( E^{tr}(x)
     \dfrac{\partial v_1}{\partial x} +
             E^{tr}(y)\dfrac{\partial v_1}{\partial y}\right )\\
  &= {\cal E}^{lr}(v_1) + \left ( E^{tr}(x)\cdot 1 + E^{tr}(y)\cdot 1\right )
  \end{split}
\end{equation}
\endgroup
Plugging Equation$~\eqref{eq:v1-unroll}$ into
Equation$~\eqref{eq:internal-node-error}$,
we obtain
the total error for $v_2$ in terms of the
local errors and incoming errors on
inputs:
\begingroup
\small
\begin{equation}
\label{eq:v2-error-full}
 {\cal E}^{tr}(v_2) = {\cal E}^{lr}(v_2) + (2x+y) E^{lr}(x)
     + x\cdot {\cal E}^{lr}(v_1) + x\cdot E^{tr}(y)
\end{equation}
\endgroup

The extent to which the error accumulated at $v_2$ impacts the output at $S$ depends
on the sensitivity of $S$ to changes in $v_2$, that is, 
	$\dfrac{\partial S}{\partial v_2} = \dfrac{\partial S}{\partial v_3}\dfrac{\partial v_3}{\partial v_2} = 3.5$

Let ${\cal E}^{tr}(S|v_2)$ denote the {\em total error component} at the output node,
$S$, due to {\em total
error accumulated} at $v_2$ (path-strength from $v_2$ to $S$); then:
\begingroup
\small
\begin{equation}
  \label{eq:v2-total-error-expr}
  \begin{split}
{\cal E}^{tr}(S|v_2) =
{\cal E}^{tr}(v_2) \dfrac{\partial S}{\partial v_2} & =
\left ({\cal E}^{lr}(v_2)
 + (2x+y)E^{lr}(x)
 + x\cdot {\cal E}^{lr}(v_1)
 + x\cdot E^{tr}(y)\right )\cdot 3.5 \\
 & = {\cal E}^{lr}(v_2)\dfrac{\partial S}{\partial v_2}
+ E^{tr}(x)\dfrac{\partial S}{\partial x} +
{\cal E}^{lr}(v_1)\dfrac{\partial S}{\partial v_1} + 
		  E^{tr}(y)\dfrac{\partial S}{\partial y}
\end{split}
\end{equation}
\endgroup

\noindent Equation$~\eqref{eq:v2-total-error-expr}$ is the symbolic form of the error
contribution from $v_2$ (and all its dependencies) to $S$.
There are two approaches to obtain the bound on
${\cal E}^{tr}(S|v_2)$:
\begin{compactenum}

\item {\em Without abstraction}, where we
feed the optimizer with the fully symbolic expression of
${\cal E}^{tr}(S|v_2) =
{\cal E}^{tr}(v_2) \dfrac{\partial S}{\partial v_2}$. This
formulation preserves all variable correlations (especially
important for non-linear expressions) and is expected
to obtain the tightest possible bounds.
However, the resulting large expressions
can often end up choking the global optimizer or excessively delay
its convergence.
%
% fully expand down to the expression
% $\bigg ({\cal E}^{lr}(v_2)
% + (2x+y)E^{lr}(x)
% + x\cdot {\cal E}^{lr}(v_1)
% + x\cdot E^{tr}(y)\bigg )\cdot 3.5$ and ... {\bf Arnab please write.}

%\item {\em With abstraction of $v_2$}, where we
% first send to the global optimizer the 
% expression ${\cal E}^{tr}(v_2)$, obtaining its bound (say $B$),
% and then introduce the free variable $FV_2$, ascribing it
% its total error which is denoted by
% $E^{tr}(FV_2)$ in Figure~\ref{fig:overview-illustration} (right-hand side).
% That is, $E^{tr}(FV_2)$ will now be set to $B$.

\item {\em With abstraction of node $v_2$}. First
we obtain a function interval and the total error 
by solving for $v_2$ and $E^{tr}(v_2) = \max(|{\cal E}^{tr}(v_2)|)$
using a global optimizer. Now, $v_2$ can be abstracted as a
{\em free variable}, $FV_2$, with an associated error term
($E^{tr}(FV_2) = E^{tr}(v_2)$)
and an interval range as shown in RHS of 
figure~\ref{fig:overview-illustration}. Then reconstruct
${\cal E}^{tr}(S|v_2) = E^{tr}(v_2)\dfrac{\partial S}{\partial v_2}$,
and feed it to the global optimizer to obtain the concrete 
bound, $E^{tr}(S|v_2)$. This reduces the the query sizes submitted
to the optimizer.
\end{compactenum}

\noindent Note
in our example that
abstracting node $v_2$ completely removed the dependency
from $v_1$ and $x$,  and, therefore, was a good choice for abstraction.
However, consider the case of $v_1$ being abstracted.
This would result in a {\em
loss of correlation between the reconvergent paths}
merging at $v_2$ for input $x$.
This can result in 
looser error bounds,
especially when error cancellations occur in the merging paths.
Error cancellations are important to include
in floating-point error analysis in order to obtain tight error bounds.

Also, to find the total error at $v_2$,
we unrolled its operand's error term,
$E^{tr}(v_1)$ as well.
Instead, if we only tracked the local error term at each node and scaled
them by their respective propagation factors (``path strengths'') to the output,
we end up with the same expression as in the last part of
equation$~\eqref{eq:v2-total-error-expr}$.
In contrast to existing tools, {\em \satire identifies this balance in finding
 designated cut-points where we can solve internal nodes for total error
 terms and abstract them, removing their child dependencies.}
This process is repeated until the AST reaches a size
that is easily managed using direct solve (unrolling all to local error terms;
see Figure~\ref{fig:gradual-abstraction}):

\section{Error Analysis}
\label{sec:error-analysis}
We now define the first-order error analysis underlying \satire
ending with a soundness claim.
Let the expression to be analyzed be presented as a straight-line
program with one output being $s$ (other outputs are similarly treated): 

\[
	s = ({\bf x}, s_1, s_2, \dots, s_n)
\]

Here $\bf{x}$ is the set
of $m$ incoming inputs, $\{x_1,x_2,\dots, x_m\}$ with $m$ incoming
error quantities, $\{e_{x_1}, e_{x_2}, \dots, e_{x_m} \}$.
Here, we think of $s_i$ as having been
computed at the $i$th step of the overall computation,
where $f_i$ are the operators supported:

\begin{equation}
\label{eq:iseq}
	s_i = f_i({\bf x}, s_1, s_2,\dots, s_{i-1})
\end{equation}

Equation$~\eqref{eq:iseq}$ represents the causality
of the execution order, that is, the operator $f_i$
being computed in the $i$th step can involve as its inputs
the primary inputs, ${\bf x}$, and any of the intermediary
computed results leading upto the $i$th stage.
When such a computation is performed in finite precision,
the operators $f_i$ are replaced by their floating-point counterparts
denoted as $\tilde{f_i}$. In addition to generating a
round-off error component at the operator site, $\tilde{f_i}$
also propagates the incoming errors through each operand.
Thus, the $i$th step of the algorithm with floating
point operators is specified as

\begin{equation}
\begin{split}
	\tilde{s_i} &= \tilde{f_i}(\tilde{{\bf x}}, \tilde{s_1}, \tilde{s_2}, \dots, \tilde{s_{i-1}}) \\
	&= \tilde{f_i}(x_1+e_{x_1},\dots,x_m+e_{x_m}, s_1 + e_{s_1}, s_2 + e_{s_2}, \dots, s_{i-1} + e_{s_{i-1}}) \\
\end{split}
\end{equation}

Here, every $e_{s_j}$ denotes the total error flowing in to the operator site due to
the accumulation at $s_j$, that is $e_{s_j} = {\cal E}^{tr}(s_j)$.
Each $\tilde{f_i}$ generates
as round-off error according to the equation$~\eqref{eq:opbound}$ , thus relating
the real and floating point evaluation as

\begin{equation}
\begin{split}
	\tilde{s_i} = f_i(x_1+e_{x_1},\dots,x_m+e_{x_m}, s_1 + e_{s_1}, \dots, s_{i-1} + e_{s_{i-1}})( 1 + \delta_i);\quad |\delta_i| \leq \bf{u}
\end{split}
\end{equation}

\noindent If $s_n$ is the final step
of the algorithm, then one could write its corresponding
floating point $\tilde{s_n}$ as

\[
	\tilde{s_n} = f_n(x_1+e_{x_1},\dots,x_m+e_{x_m}, s_1 + e_{s_1}, \dots, s_{i-1} + e_{s_{n-1}})( 1 + \delta_n);\quad |\delta_n| \leq \bf{u}
\]

A Taylor series expansion of the $f_n$ term on the rhs leads to the following 

\begin{equation}
\label{eq:taylor-expansion}
 \begin{split}
   \tilde{s_n} &= f_n(x_1+e_{x_1},\dots,x_m+e_{x_m}, s_1 + e_{s_1}, \dots, s_{i-1} + e_{s_{n-1}})( 1 + \delta_n);\quad |\delta_n| \leq \bf{u} \\
   &= \bigg ( f_n({\bf x}, s_1, \dots, s_{n-1}) + 
              \sum_{j=1}^{n-1} \dfrac{\partial s_n}{\partial s_j}e_{s_j}  +
              \sum_{j=1}^{m} \dfrac{\partial s_n}{\partial x_j}e_{x_j} 
			  \bigg ) (1 + \delta_n) + O({\bf u}^2) \\
   &= f_n({\bf x}, s_1, \dots, s_{n-1}) + f_n({\bf x}, s_1, \dots, s_{n-1})\delta_n +
              \sum_{j=1}^{n-1} \dfrac{\partial s_n}{\partial s_j}e_{s_j}  +
              \sum_{j=1}^{m} \dfrac{\partial s_n}{\partial x_j}e_{x_j} +      O({\bf u}^2) \\
   &= s_n + s_n\delta_n +
              \sum_{j=1}^{n-1} \dfrac{\partial s_n}{\partial s_j}e_{s_j}  +
              \sum_{j=1}^{m} \dfrac{\partial s_n}{\partial x_j}e_{x_j} +      O({\bf u}^2) \\
 \end{split}
\end{equation}

\satire's goal is to bound the absolute error on the output $s_n$, that is
$e_{s_n} = {\cal E}^{tr}(s_n) = |s_n - \tilde{s_n}|$. The term
$s_n\delta_n$ is the round-off error
generator at the $f_n$ operator site representing the local error
term, ${\cal E}^{lr}(s_n)$. Thus the bound on the absolute error
can be evaluated as

\begin{equation}
\label{eq:total-final-form}
\begin{split}
E^{tr}(s_n) &\leq \max(|{\cal E}^{tr}(s_n)|)  \\
%{\cal E}^{tr}(s_n) &= e_{s_n} = |s_n - \tilde{s_n}| \leq \max \bigg (
 			&\leq \max \bigg (
						|s_n \delta_n| + |\sum_{j=1}^{n-1} \dfrac{\partial s_n}{\partial s_j}e_{s_j}| +
						|\sum_{j=1}^{m} \dfrac{\partial s_n}{\partial x_j}e_{x_j} | 
					\bigg ) + O({\bf u}^2) \\
	 &\leq \max \bigg (
|{\cal E}^{lr}(s_n)| + |\sum_{j=1}^{n-1} \dfrac{\partial s_n}{\partial s_j}
      {\cal E}^{tr}(s_j)| +
	|\sum_{j=1}^{m} \dfrac{\partial s_n}{\partial x_j}{\cal E}^{tr}(x_j) | 
					\bigg ) + O({\bf u}^2) \\
\end{split}
\end{equation}

\noindent Equation$~\eqref{eq:total-final-form}$ makes three necessary distinctions as follows.

\noindent$\bullet$
 The second order error terms contain
 $\bf{u}^2$ which is an extremely small quantity
 and becomes relevant to precision
 only when $n{\bf u} \approx 1$.
 Given that ${\bf u} = 2^{-53}$, $n$ must approach $2^{53}$.
 We perform abstractions well before reaching this point.
 
%  %
%  For small operator counts we tested with FPTaylor |or second order
%  errors showing neglible impact on error bounds,
%   hence we ignore them from the analysis.

\noindent$\bullet$
 Error expressions such as
 ${\cal E}^{lr}$ and
 ${\cal E}^{tr}$ can coexits in two forms in \satire:
  (1) as a symbolic expression,
 and (2) as a numeric interval.
%
%When present as a symbolic expression,
%we have the `absolute' quantifiers
% applied
% on the symbolic terms as shown in Equation~\eqref{eq:total-final-form}.
 %
 Ultimately, when we do the abstractions or solve for the final output,
  the expressions are maximized to
  obtain the maximum error bound.

\noindent$\bullet$
 While the
 right-hand 
 side of Equation~\eqref{eq:total-final-form}
 involves additional total error terms, these
 can be further unrolled to obtain direct dependencies
 to the local error terms down the hierarchy and/or
 inputs and abstracted nodes.

%\noindent$\bullet$
%  The analysis shows the dependence of the output directly on all possible data points
%   causally allowed.
%   %
%   However, function $f_n$ defines the actual arity of $s_n$'s computation.
%   %
%   For example, $f_n$ might be just a multiplication between $s_{n-5}$ and $s_{n-1}$, in which case
%   all the other direct dependence terms in $~\eqref{eq:total-final-form}$ evaluates to 0.

%% Introduce the global optimization and canonicalization here

%\paragraph{\textbf{abstraction guided analysis \:}}

% Existing tools suffer from the
% drawback of having to choose between the ability to provide tight guarantees or
% handle large expression sizes. Building the symbolic taylor forms in FPTaylor
% introduces significant cost as the expressions begin to explode in size when
% stepping beyond small expressions. This is because at every node an {\em error variable}
% is introduced and its propagation is tracked symbolically in the formward path.
% Obtaining the full error expression is possible only after each error variable completes 
% its traversal to the output node.

\noindent As mentioned in \S\ref{sec:contributions} (and detailed now),
\satire follows an {\em incremental}
and {\em decoupled} strategy toward error analysis:
%
%introducing two separate passes in addition to incremental heuristic based abstraction
%
\begin{compactitem} 
\item A \textbf{forward pass} that assigns a local symbolic error
   term at each operator site, ${\cal E}^{lr}$.
   
\item A \textbf{backward pass} that uses {\em symbolic
  algorithmic differentiation} to derive the propagation factors (path strengths
   or derivatives) from
   each operator site to the output nodes.
 
\item \textbf{Incremental abstractions} of nodes
 as detailed in \S\ref{sec:heuristics}.
\end{compactitem}

\noindent Due to the use of abstractions,
 expressions such as ${\cal E}^{tr}(s_n)$ are comprised of
 both symbolic (``${\cal E}$'') and ground (``$E$'') representations.
More specifically, suppose $\alpha$ denotes the set of nodes that have been abstracted to
numeric (ground)
error bounds,
and $\beta = n-\alpha$ are the remaining nodes that are preserved symbolically.
Equation~\eqref{eq:total-final-form} can
be viewed as a combination of both ilks, as in:

\begin{equation}
\begin{split}
E^{tr}(s_n) &\leq \max (|\mathcal{E}^{lr}(s_n)| +
	                      |\underset{s_j \in \beta}{\sum} \dfrac{\partial s_n}{\partial s_j}\mathcal{E}^{tr}(s_j)| +
	                      |\underset{s_j \in \alpha}{\sum} \dfrac{\partial s_n}{\partial s_j}{E}^{tr}(s_i)| + \sum_{j=1}^{m} |\dfrac{\partial s_n}{\partial x_j}{E}^{tr}(x_j) |)
\end{split}
\end{equation}

%Note here that we do not separately solve for the symbolic derivative intervals, but they
%are solved in conjunction with the error expressions to obtain the net effect of each compute
%on the final bound.

\noindent Global optimizers are capable of entertaining such mixed ``concolic''
expressions.

\paragraph{\bf Soundness: \satire's error bounds are conservative:\/} This claim
is easily proved based on the following observations.
\satire's error analysis is decoupled into (1)~computing the local error symbolically, and
 (2)~multiplying it with the {\em sum of all path strengths}, again symbolically.
 Computation of the local error is based on symbolic execution to which
 the accepted error model~\cite{harrison-hol99}
 in Equation~\ref{eq:opbound}  is applied.
 The method of adding the {\em generated} errors and {\em propagating} the errors
 is justified to be sound in~\cite{rosa,higham} (also the beginning
 of \S\ref{sec:illustrative-example}).
 Computing the path strengths using {\em reverse-mode symbolic differentiation}
 is again known to be sound~\cite{automatic-diff-hovland}.
 \satire implements its own symbolic automatic differentiation.
 Last but not least, the abstractions introduced in \satire are also
 conservative in that new free variables are introduced in lieu
 of existing expression subgraphs.
 These free variables do not have any correlations with the other inputs.
 Their ground value intervals are also (recursively) computed using
 the same error analysis, and by induction (since these are subexpressions)
 their soundness follows.

%--end

%-----------------------------------------------------------------
%\section{Rigorous Abstraction-guided Error Analysis}
%\label{sec:absguided}
%\input{absguided.tex}

\section{Heuristics for Abstraction, Global Optimization}
\label{sec:heuristics}

% heuristics for abs

\paragraph{Heuristics for Abstraction:\/}
Heuristics that trigger abstraction
are based on:
(1)~relative depth (RD), measuring how close the node
is to the output (the closer it is, the less the path strength to the output);
(2)~a cost function $\phi(opCount)$,
measuring the size of the symbolic expression
and the number of free variables in it (these require increased
optimizer-time); and (3)~fanout ($fanout$) from each operator site:
large fanouts are indicative of a larger dependency footprint.
While abstracting at such sites can hurt error cancellation
due to path correlations, it helps handle heavily
interconnected networks such as
stencil-type applications or neural networks.
The depth heuristics are inspired by Shannon's information theory~\cite{Shannon1948}.
If $D$ is the total AST depth and $d_i$ is the
depth of the $i$th node, its relative depth and the overall cost function is given by:
%\[
%	\hbox{depth info}(i) = -\dfrac{d_i}{D}\log_2(\dfrac{d_i}{D})
%\]
%
%Our overall cost function is:
\begin{equation}
	\label{eq:cost}
	\begin{split}
	\hbox{depth info}(i) &= -\dfrac{d_i}{D}\log_2(\dfrac{d_i}{D}) \\
	\hbox{cost info}(i)  &=  \hbox{depth info}(i) \times (\# opCount(i))\times (\# fanout(i))
	\end{split}
\end{equation}

Additional knobs include:
(1)~a fixed depth at which an abstraction is forced (ignoring the cost function);
(2)~a [minDepth, maxDepth] window within which
\satire iteratively performs the abstraction and rebuilds ASTs.
In case the depth of the rebuilt AST is below the specified minDepth,
\satire computes the remaining expression directly (``direct-solve''
in Figure~\ref{fig:gradual-abstraction}).

\paragraph{Heuristics to Improve Global Optimization, Canonicalization:\/}
\satire uses Gelpia~\cite{gelpia-github} as the backend global optimizer.
It implements an interval branch-and-bound algorithm ({\em IBBA}~\cite{alliot2012finding})
to obtain a tight interval containing the optima when solving for the
error expressions.
Essentially, when solving
for an $n-variable$ expression, it searches
an $n$-dimensional box of intervals defined on the inputs.
IBBA produces queries by repeatedly 
dividing this $n$-dimensional box into smaller box (interval) queries,
where each query comprises of the symbolic expression and
the subdivided interval box.
These queries are fed to an interval library (Gaol)
that produces an output bound for that interval.
The final output is the best fit $n$-dimensional box that produces the tightest error bound
within a given tolerance and a limit on iterations to constrain the search algorithm.

Note that the optimizer works on real-valued expressions and must obtain a
bound that contains the optimum.
As pointed out in \S\ref{sec:tight-bounds},
Gaol can automatically canonicalize simple expressions
such as $x\cdot x\cdot x$.
However, as expressions become more complex with non-linear terms and multiple
variables, the interval subdivision process gets further bottlenecked, producing
loose intervals.
Take, for example, the `direct quadrature moments method' (DQMOM) benchmark 
captured in Equation$~\eqref{eq:dqmom}$,
where $m_i \in [-1.0,1.0]$ and each of $w_i$ and $a_i$ belongs to $[0.00001, 1.0]$:
\vspace{-0.5ex}
\begingroup
\small
\begin{equation}
\label{eq:dqmom}
\begin{split}
	r =& (0.0 + ((((w_2 * (0.0 - m_2)) * (-3.0 * ((1.0 * (a_2/w_2)) * (a_2/w_2)))) * 1.0) \\ + 
       & ((((w_1 * (0.0 - m_1)) * (-3.0 *((1.0 * (a_1/w_1)) * (a_1/w_1)))) * 1.0) \\ + 
       & ((((w_0 * (0.0 - m_0)) * (-3.0 * ((1.0 * (a_0/w_0)) * (a_0/w_0)))) * 1.0) + 0.0))))
\end{split}
\end{equation}
\endgroup

When fed to Gelpia, the unsimplified expression for $r$ in DQMOM generates the
interval {\bf [-9.0E+10, 9.0E+10]}.
However, the canonicalized expression (automatically
done by {\tt SimEngine}\footnotemark for us), is
\begingroup
\small
\[
	r = 3.0*(a_0^2)*m_0/w_0 + 3.0*(a_1^2)*m_1/w_1 + 3.0*(a_2^2)*m_2/w_2
\]
\endgroup
and this reduces the occurrence of $w_i$ and $a_i$ instances.
\footnotetext{Our use of {\tt SimEngine} is to obtain canonicalized forms of expressions
with respect to variable occurrences, and also expression simplifications. Any other engine that
has these capabilities may be used in \satire.}
This yields an interval bound
{\bf [-9.0E+05, 9.0E+05]}, which is 5 orders of magnitude tighter.
As a result, the error bound on DQMOM is
{\bf  6.9E-05} using FPTaylor and {\bf 9.6E-10} using \satire
(Table~\ref{tab:small-comparison}).
These canonicalizations plus steps to prevent canonicalization
blocking (\S\ref{sec:tight-bounds}) are largely responsible for
\satire's superior overall bounds.

%We explored further into
%FPTaylor's symbolic taylor forms and used {\em canonicalization} on them before feeding it to 
%the optimizer. This required removing the `improved rounding model' quantifiers from these taylor
%forms since they bottleneck the algebraic simplification process. This simplification leads to
%the similar order of bounds($\approx 9.6E-10)$ obtained by \satire thus stressing the fact
%that consistently performing expression simplifications before feeding them to the optimizer 
%can improve the results considerably.

%-----------------------------------------------------------------
\section{Evaluation}
\label{sec:eval}

% eval

We first evaluate \satire on
FPTaylor's  benchmarks
(also provided at~\cite{fpbench}), 
which includes a wide range of examples such as polynomials approximations
and non-linear expressions, all of which consist of a few dozen operators.
We then study \satire on a
set of fresh benchmarks comprised of much larger expressions
going up to nearly 200k operators (obtained by unrolling loops from
specific implementations of these computations).
One important example studied
is FFT by separating the complex evaluation
into real and imaginary datapaths (e.g., as done by
researchers who implement FFT circuits~\cite{schwarzlander-fft}).
We also study the Lorenz equation system, followed by
a study of relative-error profiles exhibited by FDTD.

\paragraph{Experimental Setup: }
\satire is compatible with Python3, and
its symbolic engine is based on {\tt SimEngine}~\cite{simengine} (related to SymPy).
All benchmarks were executed with Python3.8.0 version on a
dual 14-core Intel Xeon CPU E5-2680v4 2.60GHz CPUs system (total
28 processor cores) with 128GB of RAM.
%
%The comparative data
%for FPTaylor were generated on the same machine configurations.
%
To arrive at an objective comparison,
the core analysis algorithms were measured without any multicore parallelism
(both for FPTaylor and \satire).
The Gelpia solver does employ internal multithreading: we did not alter it
in any way when we used either FPTaylor or \satire.
All FPTaylor benchmarks used their specified data types. Larger benchmarks we introduced use
{\tt double} floating-point
type.
High-precision shadow value calculations were performed using
GCC's {\tt quadmath}.

\subsection{\textbf{Comparative Study}}

\begingroup
\small
\begin{table}[htbp]
\centering
\begin{tabular}{ccccccc}
\toprule
\multirow{2}{*}{Benchmarks} & \multicolumn{2}{c}{Execution Time(seconds)} & \multicolumn{3}{c}{Absolute Error} & \multirow{2}{*}{\shortstack[l]{Num \\OPs}}\\
\cmidrule(lr){2-3} \cmidrule(lr){4-6} 
& {\satire} & {FPTaylor} & {\satire} & {FPTaylor} & {\satire (UnOpt)} &  \\
\midrule
exp1x				&		{\bf 0.083}	& 0.60			& {\bf 6.19E-14} 	& 2.04E-13  	& {\bf 6.19E-14}	&	5 	\\
exp1x\_32			&		{\bf 0.087}	& 0.42			& {\bf 9.71E-06}*	& 0.00010   	& {\bf 9.71E-06}	&	5 	\\
carbonGas			&		{\bf 0.084}	& 0.74			& {\bf 2.07E-08}	& 3.07E-08  	& {\bf 9.71E-06}	&	21	\\
x\_by\_xy			&		{\bf 0.068}	& 0.42			& {\bf 1.19E-07}	& 8.31E-07  	& {\bf 9.71E-06}	&	4 	\\
verhulst			&		{\bf 0.066}	& 0.47			&  4.27E-16			& {\bf 3.79E-16}& 4.47E-16			&	10	\\
turbine3			&		{\bf 0.154} & 0.49			&  3.58E-14			& {\bf 3.48E-14}& 4.10E-14		 	&	23	\\
turbine2			&		{\bf 0.102} & 0.49			& {\bf 6.32E-14}	& 7.36E-14  	& 7.39E-14			&	19	\\
turbine1			&		{\bf 0.097}	& 0.50			& {\bf 4.89E-14}	& 5.29E-14  	& 5.60E-14		 	&	24	\\
triangle			&		0.785       & {\bf 0.72} 	&  7.05E-14			& {\bf 4.06E-14}& 7.08E-14		 	&	14	\\
test05\_nonlin1\_test2&		{\bf 0.063}	& 0.40			& 1.11E-16			& 1.11E-16  	& 1.38E-16		 	&	4 	\\
test05\_nonlin1\_r4	&		{\bf 0.068}	& 0.41			& {\bf 1.11E-06}	& 2.78E-06 		& 1.94E-06			&	6 	\\
test02\_sum8		&		{\bf 0.080}	& 0.41			& 7.77E-15			& {\bf 7.11E-15}& 7.77E-15			&	15	\\
test01\_sum3		&		{\bf 0.125} & 0.40			& {\bf 1.13E-06}	& 1.97E-06  	& {\bf 1.13E-06}	&	11	\\
sum					&		{\bf 0.079}	& 0.40			& {\bf 3.44E-15}	& 3.66E-15 		& {\bf 1.13E-06}	&	11	\\
sqrt\_add			&		{\bf 0.072}	& 0.42			& {\bf 1.17E-15}	& 8.86E-15 		& 5.22E-15			&	7 	\\
sqroot				&		{\bf 0.074} & 0.46			&  8.82E-16			& {\bf 6.00E-16}& 8.82E-16			&	25	\\
sphere				&		{\bf 0.069}	& 0.42			& {\bf 8.88E-15}	& 1.03E-14  	& {\bf 8.88E-15}	&	9 	\\
sine				&		{\bf 0.122} & 0.50			&  7.36E-16			& {\bf 6.21E-16}& 7.36E-16			&	16	\\
sineOrder3			&		{\bf 0.065} & 0.44			&  1.09E-15			& {\bf 9.56E-16}& 1.09E-15		 	&	12	\\
sec4\_example		&		{\bf 0.085}	& 0.43			& {\bf 9.96E-10}	& 1.83E-09  	& 1.30E-09			&	8 	\\
rigidBody2			&		{\bf 0.084}	& 0.43			&  3.97E-11			& {\bf 3.61E-11}& 3.97E-11			&	17	\\
predatorPrey		&		{\bf 0.068}	& 0.48			& {\bf 1.73E-16}	& 1.74E-16  	& 1.75E-16			&	12	\\
nonlin2				&		{\bf 0.081}	& 0.43			& {\bf 9.96E-10}	& 1.83E-09  	& 1.30E-09			&	8 	\\
nonlin1				&		{\bf 0.072}	& 0.44			& {\bf 1.14E-13}*	& 4.37E-11  	& 2.92E-11			&	4 	\\
logexp				&		{\bf 0.062}	& 0.41			& {\bf 3.33E-13}	& 8.90E-13  	& 4.98E-13			&	5 	\\
jetEngine			&		1.969  		& {\bf 0.84}	& {\bf 2.88E-08}	& 7.09E-08  	& {\bf 2.88E-08}	&	35	\\
intro\_example		&		{\bf 0.069}	& 0.40			& {\bf 1.14E-13}*	& 4.37E-11  	& 2.92E-11			&	4 	\\
i4					&		{\bf 0.073}	& 0.41			& {\bf 3.53E-07}*	& 8.91E-06  	& {\bf 3.53E-07}	&	5 	\\
hypot32				&		{\bf 0.066} & 0.40			& {\bf 8.43E-06}*	& 0.00052  		& {\bf 8.43E-06}	&	6 	\\
himmilbeau			&		{\bf 0.090}	& 0.45			& {\bf 9.86E-13}	& 1.00E-12  	& {\bf 9.86E-13}	&	13	\\
exp1x\_log			&		{\bf 0.094}	& 0.43			&  2.39E-12			& {\bf 1.38E-12}& 3.09E-12			&	6 	\\
bspline3			&		{\bf 0.139} & 0.41			& {\bf 7.40E-17}	& 7.86E-17 		& {\bf 7.40E-17}	&	7 	\\
delta4				&		{\bf 0.119} & 0.45			&  1.26E-13			& {\bf 1.21E-13}& 1.26E-13			&	25	\\
delta				&		{\bf 0.416} & 0.53			& 3.30E-12			& {\bf 2.41E-12}& 3.30E-12			&	45	\\
doppler1			&		{\bf 0.158} & 0.58			&  1.49E-12			& {\bf 4.06E-13}& 1.71E-12			&	16	\\
doppler2			&		{\bf 0.164} & 0.59			&  2.98E-10			& {\bf 1.03E-12}*&3.55E-10			&	 16	\\
doppler3			&		{\bf 0.161} & 0.57			&  2.70E-13			& {\bf 1.60E-13}& 3.09E-13			&	16	\\
hypot				&		{\bf 0.068}	& 0.41			& {\bf 3.37E-13}	& 9.72E-13  	& 6.53E-13		 	&	6 	\\
kepler0				&		{\bf 0.094} & 0.43			& 1.10E-13			& {\bf 1.09E-13}& 1.10E-13			&	22	\\
kepler1				&		{\bf 0.168} & 0.46			&  7.07E-13			& {\bf 4.90E-13}& 7.07E-13			&	29	\\
kepler2				&		{\bf 0.402} & 0.52			&  3.20E-12			& {\bf 2.53E-12}& 3.20E-12			&	43	\\
rigidBody1			&		{\bf 0.076}	& 0.41			&  3.80E-13			& {\bf 2.95E-13}& 3.80E-13			&	12	\\
test04\_dqmom9		&		{\bf 0.206} & 0.63			& {\bf 9.66E-10}*	& 6.91E-05 		& 2.99E-05			&	35	\\
test06\_sums4\_sum1	&		{\bf 0.320} & 0.40			& {\bf 2.38E-07}	& 6.26E-07  	& {\bf 2.38E-07}	&	7 	\\
test06\_sums4\_sum2	&		{\bf 0.082}	& 0.40			& {\bf 2.38E-07}	& 6.26E-07 		& {\bf 2.38E-07}	&	7 	\\
\end{tabular}
\caption{Comparison of Results ({\bf bold-face} highlights better results, and * highlights
  a difference of more than an order of magnitude)}
	\label{tab:satire-fptaylor}
\end{table}
\endgroup

Table ~\ref{tab:satire-fptaylor} presents the comparative results for
\satire and FPTaylor.
Comparison is presented for both the total execution times and the absolute
error bounds over a suite of 47 benchmarks exported from FPTaylor's suite.
We observe that while \satire excels in its ability to handle large
and complex expressions, even
for smaller benchmarks it obtains comparable and, in many cases, tighter bounds than
today's state-of-the-art error analysis tools.
In most cases, \satire and FPTaylor
generated error bounds within the same orders of magnitude.
\satire obtains even slightly tighter bounds in over 50\% of these
benchmarks while providing an average 4.5 times speed-up in execution time.
%
%Reasons underlying \satire's improvement for DQMOM were discussed
%in \S\ref{sec:tight-bounds}.
%
%Consider the case for the `Direct Quadrature
%Moments Method'(dqmom) benchmark. \{Briefly add with reference what dqmom does\}
%%
%FPTaylor reports an error bound of 6.909E-05 with its real function interval range in
%[-9.0E+10, 9.0E+10]. \satire obtains an extremely tight bound
%for dqmom with an absolute error bound of 9.66E-10 with its real function interval range
%in [-9.0E+05, 9.0E+05].
%
Using $10^6$ random simulations on the input 
interval ranges we empirically verify \satire's bound's 
with respect to a high-precision shadow value evaluation of these benchmarks. 
%
%Satire is able to extract this tightness of
%bounds by performing simplifications on the final error expressions in symbolic form
%before feeding into the optimizer. These simplification procedures are concerned with
%making a best effort attempt to keep single instances of an input multi or single variate term
%wherever possible by expanding the expression in a polynomial form and rearranging similar
%occuring terms together. In interval analysis, multiple occurences of the same variable fails
%to preserve any correlation between these terms leading to exploding the resulting intervals.
%Since our optimizer is an Interval Branch and Bound Analysis(IBBA), it is subject to the same
%difficulties with correlation. Thus applying the simplification procedure and canonicalizing
%expressions attempts to reduce the occurence of the same variable instances which in turn reduces the 
%loss of correlation leading to tighter bounds as seen in the dqmom example.
%%
%To support our hypothesis, we extract the Taylor forms generated by FPTaylor for dqmom and post-processed
%it through our simplification framework before to the optimizer. We observe that such a simplified taylor
%form generates significantly tighter bound of 9.62E-10 post simplification with a result comparable to \satire's
%prediction.
%

%\begin{tiny}
Furthermore, applying simplification procedures on FPTaylor's generated symbolic taylor forms, we show
improvement in its generated bounds. %
%A key insight from this analysis is obtaining tight error bounds primarily depends
%on two factors -- (1) the methodology employed to obtain the symbolic error expressions, be it as
%a taylor form expansion as in FPTyalor, or symbolic generate and propagate technique as in \satire, and
%(2) how efficiently the final error expression is processed when passing it through an optimizer
%or an iterative non linear SMT solver ~\cite{rosa}.
%%
%The second point highlights the differences one may see when attempts 
%are made to fine tune these expressions with knowledge of the solvers solving them in the backend which
%may significantly influence the resulting bounds as evident in the dqmom example.
%
By default, \satire{} optimizes the product of the {\em propagation factor} and {\em error term}, that is,
the ``absolute'' quantifiers are placed on the product terms. This helps to improve
the expression simplification process. For many of the benchmarks in Table~\ref{tab:satire-fptaylor}, 
\satire's result without this optimization shows weaker results, but still within the same order of magnitude as FPTaylor.
%\end{tiny}

\subsection{\textbf{Analysis of Larger Benchmarks}}

% \satire's provides the necessary scalability to
% handle large floating point expressions using
% an incremental error analysis approach. It introduces `node abstractions'
% iteratively during the analysis.
%
Table~\ref{tab:large-benchmarks} 
summarizes the aforesaid large benchmarks where
column `Full Solve' indicates the bound obtained without any 
abstractions. 
Without abstractions, \satire fails to obtain answers for
Lorenz, 2D-heat, and Gram-Schmidt (residual).
For linear examples, the derivatives (propagation factors for the error terms) are always constant
and, hence, do not impact the error expressions's complexity.
However, as expressions become non-linear (e.g., Lorenz),
the propagation factors are not constants
any more, leading to more complex error expressions.
{\em This is the  primary reason} why we can 
directly compute (``Full Solve'') a bound for FDTD with 192,000 operators
{\em without any abstractions}---but for Lorenz, with merely
307 operators, we had to use abstractions.
FPTaylor could not generate symbolic Taylor Forms for any of these expressions. 

While abstractions
help manage a large expression size, using abstractions frequently can impact the bounds.
On a positive note, frequent abstraction implies smaller
queries to the optimizer which helps to converge quicker exploring a smaller search space.
Conversely, frequent abstraction will also lead to loss of correlation since we are
replacing the symbolic expressions with concrete intervals. 
As a generic trend, while error bounds weaken if the frequency of abstractions is
increased, they remain in the same order of magnitude. 
In  examples such as FDTD, concretization of internal nodes during
abstraction introduces large correlation losses.
In the presence of cancellation terms (as in FDTD), preserving correlation is necessary
because it enables symbolic error variables to cancel out. Abstractions inhibit
this cancellation process and, hence, should be used judiciously in such examples.

%primary advantage in error analysis of floating point expressions is
%in its ability to scale to large expression sizes at a scale no other current
%toolset currently operates. 
%
%
%We achieve this performance capability due to
%two key attributes in \satire's design -- 1) Decoupling the full error expression
%into individual generate and propagate components \{cite Rosa \} thus reducing
%the toll of carrying around a large expression impacting memory bandwith,
%2) Introducing guided abstraction, that allows incrementally solving the total error
%terms locally at internal nodes and abstracting them with constant error terms instead of
%the symbolic expressions. Here, we analyze multiple large scale benchmarks and how each
%of these technques helps and impacts the obtained bounds.
%
%\adcmt{for non-linear examples like lorenz leads to large operator counts of the symbolic
%error terms when flattened out}

\begingroup
\small
 \setlength{\tabcolsep}{3pt}
\begin{table}[htbp]
\centering
\begin{tabular}{cccccccc}
\toprule
\multirow{3}*{Benchmarks}  & \multirow{3}*{\shortstack[l]{Num\\ OPs}} & \multicolumn{5}{c}{Absolute Error Bound} &
  \multirow{3}*{\shortstack[l]{Best\\ Execution\\ Time}}\\
\cmidrule(lr){3-7}
& & {Full Solve} & \multicolumn{4}{c}{Increasing frequency of Abstraction } & \\
%& &  & \tikzmark{x} & & & \tikzmark{y} & \\
& &  & \multicolumn{4}{c}{$\xrightarrow{\hspace*{2cm}}$} & \\
\midrule
FFT-1024pt 	& 43k &	 7.89E-13 	&	 {\bf 8.87E-13}	 &	 9.57E-13	 &	 9.22E-13	 &	 9.70E-13 & 90s \\
\midrule
Lorenz20(y) & 307  & NA & {\bf 2.53E-14} & 2.61E-14 & 2.61E-14 & 3.17E-14 & 78s\\
Lorenz40(y) & 607  & NA & {\bf 1.93E-13} & 2.69E-13 & 2.15E-13 &  2.16E-13 & 294s\\
Lorenz70(y) & 1057 & NA &  {\bf 1.88E-10} & 	3.70E-07 & 	1.07E-09 & 1.83E-09 & 1268s		\\
\midrule
\shortstack[l]{Scan(1024pt \\
              ,[-1,1])} & 3060 & 1.88E-12 & 1.88E-12 & 1.88E-12 & {\bf 1.88E-12} & 1.88E-12 & 5s \\
\midrule
fdtd(10)       & 6173 & {\bf 2.56E-14} & 7.56E-13 & 8.45E-13 & 5.43E-12 & 1.11E-10 & 17s \\
fdtd(64)       & 192k & {\bf 7.45E-13} & -- & -- & -- & -- & 2.6 hrs\\
1D-heat(10) &	442 & {\bf 6.10E-15} & 6.10E-15 & 6.10E-15 & 6.10E-15 & 6.10E-15 & 2.58s 			\\
1D-heat(32) &	4226 & 1.95E-14 & 1.95E-14 & 1.95E-14 & {\bf 1.95E-14} & 1.95E-14	& 49s	\\
2D-heat(32) &	270k & 	NA & {\bf 2.22E-14} & 2.22E-14 &2.22E-14 & 2.22E-14 & 5 hrs\\
\midrule
gSchmidt(r)  &	150 & NA & 1.14E-18 & 1.05E-18 & 1.05E-18 & 1.14E-18 & 12s \\			\\
gSchmidt(Q)  &	150 & 1.22E-18 & 	1.17E-18 & 1.13E-18 & 1.17E-18 & 1.13E-18	& 4s \\
\end{tabular} \\[0.5em]
\caption{Larger Benchmarks with and without abstractions}
\label{tab:large-benchmarks}
\end{table}

%\begin{tikzpicture}[overlay, remember picture]
%	\draw [->] ({pic cs:x}) -- ({pic cs:y}) ;
%\end{tikzpicture}

\endgroup

To further motivate the importance of obtained rigorous bounds for large benchmark
problems, we study FFT and Lorenz, both being critical
components of high precision applications.

%% case study for stencils
%%\input{stencil.tex}

%% case study for fft

\paragraph{\textbf{FFT:} } Fast Fourier transform (FFT) is an optimized algorithm for
discrete Fourier transform (DFT), which converts a finite sequence of sampled points
of a function into a same length sequence of an equally numbered complex-valued
frequency components.
It has a vast number of applications in signal processing and
fast multi-precision arithmetic for large polynomial and integer multiplications.
\label{page:fft}

An $N$-point FFT involves $\log_2N$ stages, each stage having a familiar butterfly structure
(see for example~\cite{hal-fft}).
We are not aware of any tool supporting floating-point error analysis over complex domains.
However, its application to fast math multi-precision libraries necessitates 
precise floating point error analysis and has been the subject of multiple
analytical studies~\cite{ramos,percival-mult,percival-fft,schwarzlander-fft}.
These analysis methods focus on
obtaining L2-norms in terms of root mean square (RMS) errors, or statistically profiled
error bounds.
Brisbarre et al.~\cite{hal-fft} followed up on the work from Percival's~\cite{percival-fft,percival-mult}
to report the best L2-norm bound till date of $\approx 30.99 ulps$.
That is, if $z$ represents the
discretized input samples, $Z$ is the exact result and $\hat{Z}$ is the computed result, 
then,
%\begin{equation}
%\label{eq:fft-l2}
 $  \dfrac{||Z - \hat{Z}||_2}{||Z||_2} \leq B \approx 30.99 {\bf u} $
%\end{equation}

To extend this result to a bound on absolute error corresponding to the
L-infinity norm,
we utilize two well-known relations: (1)~Between the L2 norms of the
input and outputs of an $N$-point FFT, i.e., $||Z||_2 \leq \sqrt{N} \cdot ||z||_2$,
and (2)~a generic relation between the L-infinity norm and L2-norm, i.e., 
$||Z||_2 \leq \sqrt{n}||Z||_\infty$.
Using these relations, the L-infinity norm on the
error(as obtained in ~\cite{hal-fft}) of the computed FFT result can be obtained as
\begingroup
\small
\begin{equation}
\label{eq:fft-inf}
\begin{split}
  ||Z - \hat{Z}||_\infty \leq B\cdot N \sqrt{2} ||z||_\infty
\end{split}
\end{equation}
\endgroup

Equations$~\eqref{eq:fft-inf}$ obtain the absolute-error bound analytically.
%
%Plugging in the best bound seen for l2-norm, $B$, in equation $~\eqref{eq:fft-inf}$
%obtains the corresponding the maximum absolute error bound analytically.
A 1024-point FFT with input  samples in the interval $[0,1]$ with
an L2-norm bound of $B \approx 30.99 {\bf u}$ obtains an absolute error bound of $78200 {\bf u}$.
For double precision data type, with ${\bf u} = 2^{-53}$, implying an error bound of
\textbf{4.98E-12}.

\satire partitions the real and imaginary parts
of the complex operations in FFT,
obtaining real expression types for the 
output variables guarded with rounding information at every compute stage.
Two separate datapaths are generated for the 
real and imaginary terms, each of which is solver individually.
%
%It then solves each of them individually.
%
Let $E_R$ and $E_I$ denote the absolute error bounds obtains by solving
the real and imaginary parts, respectively.
These solutions, on their own, provide the individual accuracy information of the
real and imaginary expressions.
Additionally, $E_T = \sqrt{E_R^2 + E_I^2}$ gives the
bound on the maximum of the total absolute error.
It is an upper bound on the
L-infinity norm.

We show that \satire obtains a tighter bound than the analytical
bound obtained by ~\cite{hal-fft} as in equation $~\eqref{eq:fft-inf}$.
We also select
the input space in the interval $[0,1]$.
The bound obtained for the real and imaginary parts are 
$E^R \leq 7.89E-13\ $ and $E^I \leq 7.67E-13$.
Thus the total error bound is 
$E_T \leq $ \textbf{1.1E-12} which is tighter  than the best analytical bound obtained
in ~\cite{hal-fft}. 
We tried different input intervals for FFT, each time obtaining a tight bound
in comparison to the analytical bounds of ~\cite{hal-fft}. However, the optimizer
faced convergence difficulties for intervals with zero crossings like $[-1,1]$.
In these cases, incrementally solving using abstraction of smaller depths allowed
us to solve the problem while still obtaining tighter bounds than ~\cite{hal-fft}.
%with rigorous information on the individual accuracies of the
%partitioned datapaths.

%% case study for prefix sum algorithms
%%\input{scan.tex}

%% case stidy for lorenz attarctor

\paragraph{\textbf{Lorenz equations:} }
Lorenz equations model thermally induced fluid convection
using three state variables $(x,y,z)$.
Here, $x$ represents the fluid velocity amplitude,
$y$ models temperature difference between top and bottom membranes,
while $z$ represents a distortion from linearity of temperature ~\cite{lorentz}.
The equations requires three additional parameters, $a=10$ called the Prandtl number,
$b=8/3$ corresponding to the wave number for the convection, and
$r$ being the Rayleigh number proportional to the
temperature difference. 
The recurrence relations obtained by discretizing the continuous 
versions of the Lorenz equations are given in equation $~\eqref{eq:lorenz-discrete}$,
where $k$ represents the previous iteration and $dt$ is the time discretization. 
\vspace*{-0.5ex}
\begingroup
\small
\begin{equation}
\label{eq:lorenz-discrete}
 \begin{split}
 	x_{k+1} &= x_k + a(x_k - y_k)dt;\quad
	y_{k+1} = y_k + (-x_kz_k + rx_k - y_k)dt \\
	%y_{k+1} &= y_k + (-x_kz_k + rx_k - y_k)dt \\
	z_{k+1} &= z_k + (x_ky_k - bz_k)dt \\
 \end{split}
\end{equation}
\endgroup
In~\cite{lorentz}, authors study the trajectories for different $r$ values 
for chosen initial conditions of $(x_1,y_1,z_1,dt)=(1.2,1.3,1.6,0.005)$.
It shows chaotic behavior for $r \geq 22.35$ and again starts approaching equilibrium once $r$
reaches close to 200.
However, for such chaotic systems, if two initial conditions differ by 
a quantity of $\delta$, the resulting difference after time $t$ shows exponential
separation in terms of $\delta \cdot e^{\lambda t}$.
This becomes a critical component when
evaluating such equations in finite precision since the round-off error accumulation
introduces a gradual $\delta$ error building up.

%The floating-point error accumulated introduces a proxy for the $\delta$ error build up,
%also introducing exponential separation.
%between the actual and computed trajectories.
%Obtaining precise error bounds for such computations thus becomes important. 
%
We focus on two aspects of the analysis: (1)~Obtaining bounds over a range
of input intervals over $(x,y,z)$,
%thus obtaining bounds for a known range
%of initial conditions
and (2)~Analyze the sensitivity of initial conditions on individual
inputs by using degenerate intervals on the other inputs.
%
%Such analysis is essential for
%nonlinear chaotic systems which can become unstable from accumulated noise due to 
%`exponential separation'.

Beyond a few iterations, the non-linearity involved in these equations
impede the application of current tools such as FPTaylor.
Using FPTaylor, it took 350 seconds to generate Taylor forms for a small 5-iteration case,
while timing out for unrolls beyond 10 iterations.
%FPTaylor took around 350 seconds to build the taylor forms for an unroll of 5 iterations
%while timing out beyond an hour for 10 iterations.
%
Additionally, the symbolic Taylor
form for 5 iterations was so large that backend optimizer could not handle.

Using \satire{}, we obtain tight bounds for as large as 70 iterations of the problem using
abstraction-guided method.
%
%In this analysis we discuss how abstraction allows us to push the
%limits of analysis significantly while also highlighting when it hits certain limits.
%beyond which it looses significance.
%
Note here the non-linearity of
these equations makes it difficult to simplify expressions beyond a certain limit.
\satire delays further canonicalization beyond an operational count larger than 8,000,
controlled by a parametric knob, $maxOpCount$, with default value of 8,000 selected 
over multiple experiments over a mix of non-linear and linear systems.

%When flattening out the Lorenz expressions, due to non-linearity, the return on effort spent for 
%reassociation is diminished.
%%
%This is due to generation of many multi-variate composite variable terms with each of $(x,y,z)$ 
%associated with different degrees. 
%
The error expressions composed of the products of forward error and reverse derivative
may reach an operator count of over 100K over just 20 iterations, choking {\tt simEngine}'s
simplification process. Using $maxOpCount$, we only allow simplification within the depth
necessary for abstraction. During the next abstraction, further simplification will
occur.
%Over say 20 iterations, the completely flattened expression may reach an operation count of over 100K
%for the error expression which is a combination of the forward error and reverse derivatives. 
%%
%To reduce the overhead of SimEngine, we do not simplify the expression nodes beyond the 8K heuristically. 
%%

%\skcmtside{Don't quite get this sentence}
%Furthermore, we require abstraction guidance here else the large expression sizes will become untenable for the optimizers. 
%
For example, for the Lorenz-20 (that is unrolled over 20 iterations), the overall AST depth is 89.
Table~\ref{tab:lorenz-table} shows the
bounds obtained for varying windows of the abstraction depth and
the corresponding execution time(exec\_time) in seconds.

%\satire has 2 knobs called the `\--mindepth' and `\--maxdepth' that informs it a range within which to perform the \ggcmtside{cut? or shrink!}
%abstraction. Suppose we provide the knob values such that $mindepth=15$ and $maxdepth=25$, hence asking 
%\satire to find an optimal abstraction depth in that range guided by the expression sizes , depth and the 
%fan-outs from each node. Suppose the abstraction depth obtained was 18. Then all nodes in the AST at a 
%bottom-up depth of 18 are abstracted  that is each of these nodes are separately solved for their total errors,
%$E^{tr}$ and then replaced with free variables with an incoming error equalling their corresponding total
%error value. To summarize, after one abstraction level,
%now we have an AST with a depth atleast 71(we say minimum because if there exists some dependencies crossing over the abstraction depth, like from depth 10 to 
%depth 25 directly, those will still exists. The AST is then rebuilt with the freshly allocated free variables + 
%any non-dangling inputs, that is , inputs still having dependencies post-abstraction. This iterative process hence
%avoids the issue of large expression sizes and operation counts by incrementally simplying the AST
%only till the required abstracion sizes and we constrain it to be within the tuned parameter of $maxOpCount$.

%Of course, the larger abstraction depth we select, the lesser loss of correlation 
%needs to be accounted for, however constrained by solving for large error expression. 
\begingroup
\small
\begin{table}[htbp]
\centering
\begin{tabular}{ccccccc}
\toprule
\multirow{3}{*}{Lorenz} & \multicolumn{6}{c}{Window of Abstraction depth : (mindepth, maxdepth)} \\
\cmidrule(lr){2-7}
& \multicolumn{2}{c}{(5,10)} & \multicolumn{2}{c}{(10,15)} & \multicolumn{2}{c}{(15,25)} \\
\cmidrule(lr){2-3} \cmidrule(lr){4-5} \cmidrule(lr){6-7}
				 & err 		& exec\_time 			 & err		& exec\_time 			& err 		& exec\_time  \\
lorenz20: x	 & 1.61E-14 & \multirow{3}{*}{284s}	 & 1.18E-14 & \multirow{3}{*}{86s} 	& 1.16E-14  & \multirow{3}{*}{76s}  \\
lorenz20: y	 & 3.17E-14	&  		 				 & 2.62E-14	&   		 			& 2.53E-14	& 	\\
lorenz20: z	 & 1.25E-14	&  		 				 & 9.75E-15 &   		 			& 9.59E-15	& 	\\
\cmidrule(lr){2-7}
lorenz40: x	 & 8.30E-14 & \multirow{3}{*}{1200s}	 & 1.10E-13	& \multirow{3}{*}{981s}	& 7.81E-14	& \multirow{3}{*}{304s}	\\
lorenz40: y	 & 2.15E-13 &     		 			 & 2.69E-13	&    		 			& 1.95E-13	& 	\\
lorenz40: z	 & 1.66E-13 &     		 			 & 2.08E-13	&    		 			& 1.49E-13	& 	\\
\end{tabular}
\caption{Error bounds for the three state variables in the Lorenz equation}
\label{tab:lorenz-table}
\end{table}
\endgroup
%Table ~\ref{tab:lorenz-table} shows the execution time and error bounds obtained
%by \satire for three different levels of abstraction. 
We perform analysis for both 20 and
40 iterations. The second state variable, $y$, shows more worst-case variation. 
Therefore, we studied the sensitivity of $y$ for a larger size of 70 iterations 
using degenerate intervals in $x$ and $y$,
obtaining a bound of 1.11E-11 as opposed to 1.9E-10 in the non-degenerate case.

%-- \subsection{\textbf{Empirical Evaluation}}         

\subsection{\textbf{Relative error profiles}}

Error bounds obtained using static analysis methods generally provide the
worst-case absolute-error bounds for a given input range.
However, absolute error is not indicative of true precision-loss:
a worst-case absolute error of,
say, $0.01$ is large when the actual value is close to 1 than when it is 100.
Relative error, obtained by normalizing the absolute error with the
true value of the variable, provides more insight into precision.
Given a real-valued expression $f$, its floating point counterpart $\tilde{f}$, 
then
$re(f) = ({|f - \tilde{f}|}/{f})$ denotes its relative error.

Using \satire{}, we can obtain a maximum bound for the numerator, that is,
$\max(|f - \tilde{f}|)$.
Tools that attempt to obtain a relative-error bound
statically do so in two primary ways: (1) divide the worst-case error by the lower bound
of the real-valued function interval, or (2) obtain a symbolic expression for $re(f)$
and solve the optimization problem.
While the prior technique leads to 
quite conservative bounds (loss of correlation between the
numerator and denominators), the latter suffers when encountering intervals
close to or containing $0$.

In practice, however, the output is only rarely close to $0$.
Our approach capitalizes on this to obtain dynamic information on the relative error,
making use of the observed runtime outputs.
More precisely, we want to quantify the numer of bits lost
using \satire{}-computed worst-case absolute error. 
For a floating-point system using $p$ precision bits and a machine
epsilon of ${\bf u}$, the number of bits lost is obtained as
$\# bits\ lost = p - \log_2(re(f)/{\bf u})$.
%\vspace{-0.5ex}
%\begingroup
%\small
%\begin{equation}
%\label{eq:bits-lost}
%	\# bits\ lost = p - \log_2(re(f)/{\bf u})
%\end{equation}
%\endgroup
%\vspace{-0.5ex}

The information we do not have here is the true value of $f$.
To this end, we use $\tilde{f}$ as a proxy for $f$ in the denominator.
Since $\tilde{f}$ is the observed value at runtime, there is no extra overhead to evaluate
it (unlike shadow value calculations).

We use FDTD as a case study here. We select FDTD because it was the worst-behaving 
benchmark when using frequent abstractions (attributable
to the large number of  cancellation terms), implying a larger impact on precision bits
than additive kernels like heat.
We analyze it in the input interval
$[0,1]$ to obtain values closer to zero, because floating-point values
are highly condensed in this region and {\em many binades are included in this interval.}

%\begin{figure}
%	\centering
%    \includegraphics[width=0.6\textwidth,height=0.4\columnwidth]{figs/fdtd-rel-error.PNG}
%  \caption{Difference in prediction of loss of precision bits using Satire and shadow value }
%  \label{fig:fdtd-rel-error}
%\end{figure}
%--
\vspace{-0.5ex}
\begin{figure*}
  \begin{minipage}{.25\textwidth}
    %\begin{equation}
	%	\label{eq:bits-lost}
	%  \# bits\ lost = p - \log_2(re(f)/{\bf u})
    %\end{equation}
    \begin{equation}
		\label{eq:compare-rel-error}
	  \begin{split}
		Q &= qsat - qshadow; \\
		qsat &= \dfrac{\max(|f - \tilde{f}|)}{|f_{dp}|} ; \\
		qshadow &= \dfrac{|f_{qp} - f_{dp}|}{|f_{qp}|}; \\
	  \end{split}
    \end{equation}
  \end{minipage}\hfill
  \begin{minipage}{.65\textwidth}
    \centering
    \includegraphics[width=0.8\columnwidth,height=0.6\columnwidth]{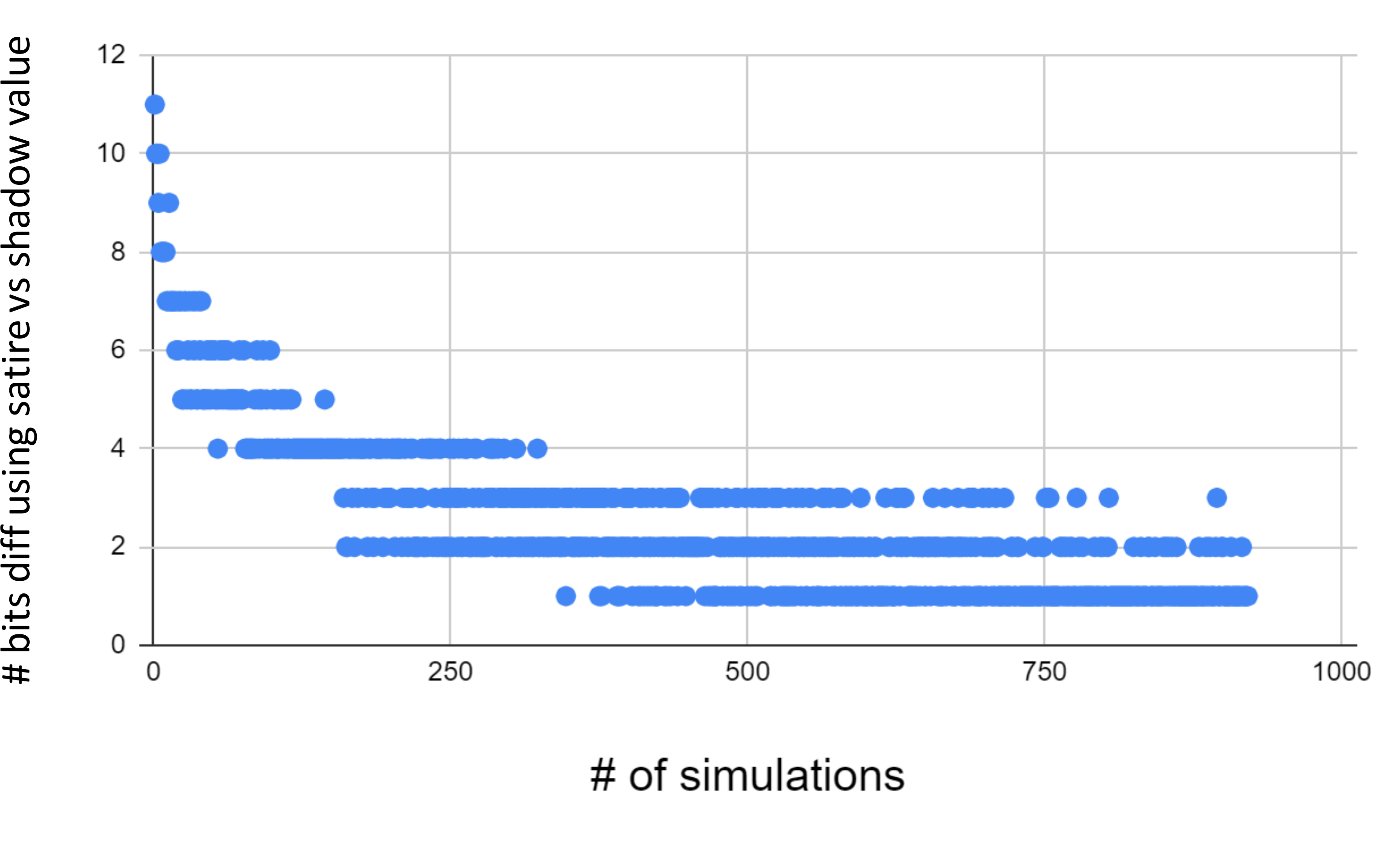}\\[-0.5em]
    \caption{\label{fig:fdtd-rel-error}
Difference in precision estimate(\# bits lost) between \satire and shadow value calculation}
    \end{minipage}
\end{figure*}

%\vspace*{-0.5ex}

Even for shadow-value evaluations,
one uses a high-precision value as a proxy for the ``true value.''
Let $f_{dp}$ and $f_{qp}$ denote the observed values when using
double and quad (high) precision, respectively.
In figure~\ref{fig:fdtd-rel-error}, the y-axis plots the 
prediction difference, $Q$, evaluated as shown in equation$~\eqref{eq:compare-rel-error}$
and the x-axis plots the simulations done for the empirical evaluation.
Here $qsat$, the numerator, is from \satire's worst case-bounds, 
the denominator is the observed value in the working precision (here, double).
For $qshadow$, the numerator is the observed difference in dynamic data between
double and quad precision values,
while its denominator is the quad precision value.
%\begingroup
%\small
%\begin{equation}
%\label{eq:compare-rel-error}
%\begin{split}
%	Q &= qsat - qshadow;\quad
%	qsat = \dfrac{\max(|f - \tilde{f}|)}{|f_{dp}|} ;\quad
%	qshadow = \dfrac{|f_{qp} - f_{dp}|}{|f_{qp}|} \\
%	%Q &= qsat - qshadow
%\end{split}
%\end{equation}
%\endgroup
%\vspace{-0.5ex}

%In $qsat$, the numerator is from \satire's worst case-bounds, 
%the denominator is the observed value in the working precision (here, double).
%%
%For $qshadow$, the numerator is the observed difference in dynamic data between
%double and quad precision values,
%while its denominator is the quad precision value.

In practice, using shadow-value calculations is a large burden for the user:
an entirely duplicated
thread has to be run in high precision, with additional necessary code changes.
Our objective is to help the user avoid all this and still allow them
to gain value through \satire{}'s analysis toward precision loss.
We aim to show that using only the observed values, we can precisely inform the 
loss in precision bits.
In other words, $Q$ should be always greater than zero 
for soundness but close to zero for tightness.
Figure~\ref{fig:fdtd-rel-error}
shows denser regions closer to zero implying the tightness of our analysis compared to
the shadow value calculation, while never crossing zero indicating a sound bound.
Thus, we believe that users can use \satire{}'s analysis to obtain a quick indication of
precision loss---at least in a relative sense.
That is, they may be able to use our methods to
{\em zoom into code regions that have higher loss and improve 
overall higher precision.}

%--

%--end

%-----------------------------------------------------------------
\section{Additional Related Work}
\label{sec:relwork}
Precision analysis plays an important role in safety-critical
systems.
For trustworthy error analysis
formally certified bounds are needed, and
some recent examples are~\cite{flover-darulova,DBLP:journals/toplas/SolovyevBBJRG19}.

Rosa and FPTaylor are the closest to our work
in their approach to extend Taylor forms rigorous
floating point error estimation. 
Rosa propagates 
errors in numeric affine form and uses SMT solvers
to obtain tight bounds.
In FPTaylor, full symbolic Taylor forms are obtained
and fed to a global optimizer, which often results
in large expressions, impeding scalability.
%
%In contrast, \satire decouples the taylor form
%into error generation and path strength derivation, as
%part of separate passes and not in one swoop.
%both being symbolic, 
%
In contrast, \satire's decoupled analysis 
achieves the same end-result albeit without
this complexity, by using a forward and backward pass 
that greatly improves overall effectiveness.

%
%Gappa~\cite{gappa} is basically an interval-based
%reasoning system, but comes with many simplification
%rules of its own, and has been embedded into
%verifiers such as~\cite{frama-c}
%and various proof-assistants~\cite{coq-fp,hollight-fp,precisa}.
%Rosa and FPTaylor are the closest to \satire in terms of approach.
%%
%Rosa extends Taylor analysis
%by computing function expressions symbolically and
%propagates errors using a numeric affine representation.
%%
%It uses SMT solvers for obtaining accurate bounds.
%
%FPTaylor tracks dependencies symbolically by introducing
%noise variables at each operator,
%propagating them symbolically along forward paths.
%%
%As pointed out earlier, this approach to Taylor-form generation
%cannot scale.
%%
%\satire's decoupled analysis achieves the same end-result albeit without
%this complexity.
%
%It decouples the symbolic Taylor form
%generation
%by avoiding propagation of error expressions in the forward path.
%%
%It resorts to a backward pass using reverse mode symbolic differentiation
%to obtain effective path strengths.

%Real2Float approaches error analysis
%based on semidefinite programming, and
%bounds higher-order error 
%similar to how it is done in FPTaylor.
%
%While second-order error analysis is important,
%the difficulty of obtaining this in meaningful settings
%has already been noted in \S\ref{sec:intro}.

%Abstraction-based methods,
%including abstract interpretation~\cite{cousot-abstract}
%and polyhedra~\cite{sound-fp-polyhedra} have been pursued.

While scalability has been previously achieved,
it required manual assistance using 
abstract interpretation~\cite{cousot-abstract}
coupled with
theorem proving (e.g., FLuctuat~\cite{fluctuat,coq-wave}).
Gappa~\cite{gappa} is basically an interval-based
reasoning system, but comes with many simplification
rules of its own, and has been embedded into
verifiers such as~\cite{frama-c}
and various proof-assistants~\cite{coq-fp,hollight-fp,precisa}.
Combined uses of static analyses~\cite{static-fp}
are popular, as demonstrated at scale in
Astr{\'e}e~\cite{cousot-astree}. 
A general abstract domain for floating-point computations is described
in~\cite{martel-semantics}.

%-eva rel error
The importance of sound techniques for relative error
estimation has been recognized by many.
A common approach for relative error estimation is to
first obtain the absolute error and then divide it by
the minimum of the function interval value.
A combined rigorous approach to relative error estimation is
presented in~\cite{eva-rel-error}.
Our approach for relative error estimation
is meant for use in a dynamic analysis setting.
It is motivated by the important
pragmatic consideration of avoiding shadow-value
computations on an existing piece of code.
We use the observed value at runtime in combination
with \satire's worst-case bound to obtain a relative
error profile that provides insights on precision loss.
Through empirical evaluation, we establish the reliability
of this approach.

%-eva cert

%--end

%-----------------------------------------------------------------
\section{Concluding Remarks}
\label{sec:conc}
We presented \satire, a tool for rigorous floating-point error
analysis that produces tight error bounds in practice.
\satire is similar to many of its predecessors,
but specifically emphasizes handling large expressions that
arise in practice by including
an information-theoretic abstraction mechanism
for scalability.
The effectiveness of \satire
has been demonstrated on practical examples including
FFT, parallel prefix sum, and stencils for
various partial differential equation (PDE) types.
Even divergent families of equations such as
the Lorenz system are included in our study.
\satire can provide insights on the loss of precision
at runtime as demonstrated on a large ill-conditioned problem.
We believe this variety and scale in an automated rigorous tool
is unique.

Our work quickly showed us that rather than merely the size of 
expressions, one must also take the {\em kinds of computations}
being analyzed.
For instance, very large expressions generated by FDTD and heat-flow
can be analyzed without using abstractions for large sizes; however,
highly non-linear systems such as Lorenz {\em require} abstractions
even for small sizes.

An important conclusion that emerges pertains to
the impact of losing variable correlations
due to abstractions: it is an important consideration
in choosing when and where to abstract.
The importance of incremental and decoupled
error expression computation is also brought out.
Given that global optimizers are workhorses in error estimation,
our studies shed light on the role of expression canonicalization.
This fits well with incremental computation, allowing global optimizer
calls to be smaller as well as can be parallelized.

Important future directions include
handling loops (enabling further scaling),
improving abstractions without increasing error bounds,
and the use of parallelism to further speed up the analysis.

%--

\clearpage

\bibliography{bib/satire,bib/ganesh-nsfmed-2019,bib/FM_dmtcp.bib}

\def \noopsort #1{}
\begin{thebibliography}{10}
\providecommand{\url}[1]{\texttt{#1}}
\providecommand{\urlprefix}{URL }
\providecommand{\doi}[1]{https://doi.org/#1}

\bibitem{alliot2012finding}
Alliot, J.M., Durand, N., Gianazza, D., Gotteland, J.B.: Finding and proving
  the optimum: Cooperative stochastic and deterministic search. In: Proceedings
  of the 20th European Conference on Artificial Intelligence (ECAI). pp.
  55--60. ACM (2012). \doi{10.3233/978-1-61499-098-7-55}

\bibitem{numstats}
Altman, M., Gill, J., McDonald, M.P.: Numerical Issues in Statistical Computing
  for the Social Scientist. John Wiley {\&} Sons, Inc. (Dec 2003).
  \doi{10.1002/0471475769}, \url{https://doi.org/10.1002/0471475769}

\bibitem{flover-darulova}
Becker, H., Zyuzin, N., Monat, R., Darulova, E., Myreen, M., Fox, A.: A
  verified certificate checker for finite-precision error bounds in coq and
  hol4. In: FMCAD. pp. 1--10 (10 2018). \doi{10.23919/FMCAD.2018.8603019}

\bibitem{automatic-diff-hovland}
Bischof, C., Buker, H., Hovland, P., Naumann, U., Utke, J. (eds.): Advances in
  Automatic Differentiation. Springer (2008), iSBN : 978-3-540-68935-5

\bibitem{boldo-2nd-order-wave}
Boldo, S., Cl{\'{e}}ment, F., Filli{\^{a}}tre, J.C., Mayero, M., Melquiond, G.,
  Weis, P.: Wave equation numerical resolution: A comprehensive mechanized
  proof of a c program. Journal of Automated Reasoning  \textbf{50}(4),
  423--456 (Aug 2012). \doi{10.1007/s10817-012-9255-4},
  \url{https://doi.org/10.1007/s10817-012-9255-4}

\bibitem{coq-wave}
Boldo, S., Cl{\'e}ment, F., Filli{\^{a}}tre, J.C., Mayero, M., Melquiond, G.,
  Weis, P.: Wave equation numerical resolution: A comprehensive mechanized
  proof of a {C} program. Journal of Automated Reasoning (JAR)  \textbf{50}(4),
   423--456 (2013). \doi{10.1007/s10817-012-9255-4}

\bibitem{percival-mult}
Brent, R.P., Percival, C., Zimmermann, P.: Error bounds on complex
  floating-point multiplication. Math. Comput.  \textbf{76},  1469--1481 (2007)

\bibitem{hal-fft}
Brisebarre, N., Joldes, M., Muller, J.M., Nane{\c s}, A.M., Picot, J.: {Error
  analysis of some operations involved in the Cooley-Tukey Fast Fourier
  Transform}. {ACM Transactions on Mathematical Software} pp. 1--34 (2019)

\bibitem{fptuner}
Chiang, W.F., Baranowski, M., Briggs, I., Solovyev, A., Gopalakrishnan, G.,
  Rakamariundefined, Z.: Rigorous floating-point mixed-precision tuning.
  SIGPLAN Not.  \textbf{52}(1),  300–315 (Jan 2017).
  \doi{10.1145/3093333.3009846}, \url{https://doi.org/10.1145/3093333.3009846}

\bibitem{cousot-abstract}
Cousot, P., Cousot, R.: Abstract interpretation: A unified lattice model for
  static analysis of programs by construction or approximation of fixpoints.
  In: Proceedings of the 4th ACM SIGACT-SIGPLAN Symposium on Principles of
  Programming Languages (POPL). pp. 238--252. ACM (1977)

\bibitem{cousot-astree}
Cousot, P., Cousot, R., Feret, J., Mauborgne, L., Min\'e, A., Monniaux, D.,
  Rival, X.: The {ASTR\'EE} analyser. In: Proceedings of the 14th European
  Symposium on Programming Languages and Systems (ESOP). Lecture Notes in
  Computer Science, vol.~3444, pp. 21--30. Springer (2005)

\bibitem{rosa}
Darulova, E., Kuncak, V.: Sound compilation of reals. In: Proceedings of the
  41st ACM SIGPLAN-SIGACT Symposium on Principles of Programming Languages
  (POPL). pp. 235--248. ACM (2014)

\bibitem{gappa}
Daumas, M., Melquiond, G.: {Certification of Bounds on Expressions Involving
  Rounded Operators}. ACM Transactions on Mathematical Software
  \textbf{37}(1),  1--20 (2010)

\bibitem{fluctuat}
Delmas, D., Goubault, E., Putot, S., Souyris, J., Tekkal, K., V{\'e}drine, F.:
  {Towards an Industrial Use of {FLUCTUAT} on Safety-Critical Avionics
  Software}. In: Formal Methods for Industrial Critical Systems, {FMICS} 2009,
  Lecture Notes in Computer Science, vol.~5825, pp. 53--69. Springer Berlin
  Heidelberg (2009). \doi{10.1007/978-3-642-04570-7\_6}

\bibitem{verificarlo}
Denis, C., de~Oliveira~Castro, P., Petit, E.: Verificarlo: Checking floating
  point accuracy through monte carlo arithmetic. In: Montuschi, P., Schulte,
  M.J., Hormigo, J., Oberman, S.F., Revol, N. (eds.) 23nd {IEEE} Symposium on
  Computer Arithmetic, {ARITH} 2016, Silicon Valley, CA, USA, July 10-13, 2016.
  pp. 55--62. {IEEE} Computer Society (2016). \doi{10.1109/ARITH.2016.31},
  \url{https://doi.org/10.1109/ARITH.2016.31}

\bibitem{fpbench}
{FPBENCH}, benchmarks, compilers and standards for the floating point research
  community. \url{http://fpbench.org/benchmarks.html}

\bibitem{frama-c}
{{Frama-C} Software Analyzers}. \url{http://frama-c.com/index.html} (2017)

\bibitem{gelpia-github}
Gelpia: A global optimizer for real functions (2017),
  \url{https://github.com/soarlab/gelpia}

\bibitem{munoz-gappa-pvs}
Goodloe, A.E., Mu{\~{n}}oz, C., Kirchner, F., Correnson, L.: Verification of
  numerical programs: From real numbers to floating point numbers. In: Brat,
  G., Rungta, N., Venet, A. (eds.) NASA Formal Methods. pp. 441--446. Springer
  Berlin Heidelberg, Berlin, Heidelberg (2013)

\bibitem{static-fp}
Goubault, E., Putot, S.: {Static Analysis of Finite Precision Computations}.
  In: International Workshop on Verification, Model Checking, and Abstract
  Interpretation, VMCAI 2011, Lecture Notes in Computer Science, vol.~6538, pp.
  232--247. Springer Berlin Heidelberg (2011)

\bibitem{hifptuner}
Guo, H., Rubio-González, C.: Exploiting community structure for floating-point
  precision tuning. In: ISSTA. pp. 333--343 (07 2018).
  \doi{10.1145/3213846.3213862}

\bibitem{harrison-hol99}
Harrison, J.: A machine-checked theory of floating point arithmetic. In:
  Bertot, Y., Dowek, G., Hirschowitz, A., Paulin, C., Th{\'e}ry, L. (eds.)
  Theorem Proving in Higher Order Logics: 12th International Conference,
  TPHOLs'99. Lecture Notes in Computer Science, vol.~1690, pp. 113--130.
  Springer-Verlag, Nice, France (1999)

\bibitem{hollight-fp}
Harrison, J.: {Floating-Point Verification Using Theorem Proving}. In: SFM
  2006, Lecture Notes in Computer Science, vol.~3965, pp. 211--242. Springer
  Berlin Heidelberg (2006)

\bibitem{higham}
Higham, N.J.: Accuracy and Stability of Numerical Algorithms. Society for
  Industrial and Applied Mathematics, second edn. (2002).
  \doi{10.1137/1.9780898718027},
  \url{https://epubs.siam.org/doi/abs/10.1137/1.9780898718027}

\bibitem{eva-rel-error}
Izycheva, A., Darulova, E.: On sound relative error bounds for floating-point
  arithmetic. In: Proceedings of the 17th Conference on Formal Methods in
  Computer-Aided Design. p. 15–22. FMCAD ’17, FMCAD Inc, Austin, Texas
  (2017)

\bibitem{DBLP:conf/sas/JacqueminPV18}
Jacquemin, M., Putot, S., V{\'{e}}drine, F.: A reduced product of absolute and
  relative error bounds for floating-point analysis. In: Podelski, A. (ed.)
  Static Analysis - 25th International Symposium, {SAS} 2018, Freiburg,
  Germany, August 29-31, 2018, Proceedings. Lecture Notes in Computer Science,
  vol. 11002, pp. 223--242. Springer (2018).
  \doi{10.1007/978-3-319-99725-4\_15},
  \url{https://doi.org/10.1007/978-3-319-99725-4\_15}

\bibitem{schwarzlander-fft}
Jr, E., Saleh, H.: Fft implementation with fused floating-point operations.
  Computers, IEEE Transactions on  \textbf{61},  284 -- 288 (03 2012).
  \doi{10.1109/TC.2010.271}

\bibitem{craft}
Lam, M.O., Hollingsworth, J.K.: Fine-grained floating-point precision analysis.
  The International Journal of High Performance Computing Applications
  \textbf{32}(2),  231--245 (Jun 2016). \doi{10.1177/1094342016652462},
  \url{https://doi.org/10.1177/1094342016652462}

\bibitem{aiken-math-dot-h}
Lee, W., Sharma, R., Aiken, A.: On automatically proving the correctness of
  math.h implementations. {PACMPL}  \textbf{2}({POPL}),  47:1--47:32 (2018).
  \doi{10.1145/3158135}, \url{https://doi.org/10.1145/3158135}

\bibitem{lorentz}
Liang, J., Song, W.: Difference equation of lorenz system. International
  Journal of Pure and Apllied Mathematics  \textbf{83} (02 2013).
  \doi{10.12732/ijpam.v83i1.9}

\bibitem{MagronCD15}
Magron, V., Constantinides, G., Donaldson, A.: Certified roundoff error bounds
  using semidefinite programming. ACM Transactions on Mathematical Software
  \textbf{43}(4),  34:1--34:31 (Jan 2017). \doi{10.1145/3015465},
  \url{http://doi.acm.org/10.1145/3015465}

\bibitem{martel-semantics}
Martel, M.: Semantics of roundoff error propagation in finite precision
  calculations. Higher Order Symbolic Computation  \textbf{19}(1),  7--30
  (2006). \doi{10.1007/s10990-006-8608-2},
  \url{http://dx.doi.org/10.1007/s10990-006-8608-2}

\bibitem{economic}
Mccullough, B.D., Vinod, H.: The numerical reliability of econometric software.
  Journal of Economic Literature  \textbf{37},  633--665 (02 1999).
  \doi{10.1257/jel.37.2.633}

\bibitem{coq-fp}
Melquiond, G.: {Floating-Point Arithmetic in the Coq System}. Information and
  Computation  \textbf{216},  14--23 (2012). \doi{10.1016/j.ic.2011.09.005},
  \url{http://dx.doi.org/10.1016/j.ic.2011.09.005}

\bibitem{adapt}
Menon, H., Lam, M.O., Osei-Kuffuor, D., Schordan, M., Lloyd, S., Mohror, K.,
  Hittinger, J.: Adapt: Algorithmic differentiation applied to floating-point
  precision tuning. In: Proceedings of the International Conference for High
  Performance Computing, Networking, Storage, and Analysis. SC ’18, IEEE
  Press (2018)

\bibitem{munoz-winding}
Moscato, M.M., Titolo, L., Feli{\'u}, M.A., Mu{\~{n}}oz, C.A.: Provably correct
  floating-point implementation of a point-in-polygon algorithm. In: ter Beek,
  M.H., McIver, A., Oliveira, J.N. (eds.) Formal Methods -- The Next 30 Years.
  pp. 21--37. Springer International Publishing, Cham (2019)

\bibitem{munoz-gappa}
Narkawicz, A., Mu{\~{n}}oz, C., Dutle, A.: Formally-verified decision
  procedures for univariate polynomial computation based on sturm's and
  tarski's theorems. Journal of Automated Reasoning  \textbf{54}(4),  285--326
  (Feb 2015). \doi{10.1007/s10817-015-9320-x},
  \url{https://doi.org/10.1007/s10817-015-9320-x}

\bibitem{herbie}
Panchekha, P., Sanchez-Stern, A., Wilcox, J.R., Tatlock, Z.: Automatically
  improving accuracy for floating point expressions. In: Proceedings of the
  36th ACM SIGPLAN Conference on Programming Language Design and
  Implementation, {PLDI} 2015. pp. 1--11. ACM (2015).
  \doi{10.1145/2737924.2737959},
  \url{http://doi.acm.org/10.1145/2737924.2737959}

\bibitem{patriot-failure}
Failure of the Patriot Missle due to Floating-Point Error Accumulation.
  \url{http://www.ima.umn.edu/~arnold/455.f96/disasters.html}

\bibitem{percival-fft}
Percival, C.: Rapid multiplication modulo the sum and difference of highly
  composite numbers. Math. Comput.  \textbf{72}(241),  387395 (Jan 2003).
  \doi{10.1090/S0025-5718-02-01419-9},
  \url{https://doi.org/10.1090/S0025-5718-02-01419-9}

\bibitem{ramos}
Ramos, G.: Roundoff error analysis of the fast fourier transform. Mathematics
  of Computation - Math. Comput.  \textbf{25},  757--757 (10 1971).
  \doi{10.1090/S0025-5718-1971-0300488-0}

\bibitem{sc13-precimonious}
Rubio-Gonz\'{a}lez, C., Nguyen, C., Nguyen, H.D., Demmel, J., Kahan, W., Sen,
  K., Bailey, D.H., Iancu, C., Hough, D.: Precimonious: Tuning assistant for
  floating-point precision. In: Supercomputing (SC). pp. 27:1--27:12 (2013),
  \url{https://github.com/corvette-berkeley/precimonious}

\bibitem{rocco-solver-nfm-2019}
Salvia, R., Titolo, L., Feli{\'{u}}, M.A., Moscato, M.M., Mu{\~{n}}oz, C.A.,
  Rakamaric, Z.: A mixed real and floating-point solver. In: Badger, J.M.,
  Rozier, K.Y. (eds.) {NASA} Formal Methods - 11th International Symposium,
  {NFM} 2019, Houston, TX, USA, May 7-9, 2019, Proceedings. Lecture Notes in
  Computer Science, vol. 11460, pp. 363--370. Springer (2019).
  \doi{10.1007/978-3-030-20652-9\_25},
  \url{https://doi.org/10.1007/978-3-030-20652-9\_25}

\bibitem{stoke-fp}
Schkufza, E., Sharma, R., Aiken, A.: {Stochastic Optimization of Floating-point
  Programs with Tunable Precision}. In: PLDI 2014. pp. 53--64. PLDI '14, ACM
  (2014)

\bibitem{Shannon1948}
Shannon, C.E.: A mathematical theory of communication. The Bell System
  Technical Journal  \textbf{27}(3),  379--423 (7 1948).
  \doi{10.1002/j.1538-7305.1948.tb01338.x},
  \url{https://ieeexplore.ieee.org/document/6773024/}

\bibitem{simengine}
Simengine. \url{https://www.ensoftcorp.com/simengine/}

\bibitem{DBLP:journals/toplas/SolovyevBBJRG19}
Solovyev, A., Baranowski, M.S., Briggs, I., Jacobsen, C., Rakamaric, Z.,
  Gopalakrishnan, G.: Rigorous estimation of floating-point round-off errors
  with symbolic taylor expansions. {ACM} Trans. Program. Lang. Syst.
  \textbf{41}(1),  2:1--2:39 (2019). \doi{10.1145/3230733},
  \url{https://doi.org/10.1145/3230733}

\bibitem{affine}
Stolfi, J., de~Figueiredo, L.H.: {An Introduction to Affine Arithmetic}. TEMA
  Trends in Applied and Computational Mathematics  \textbf{4}(3),  297--312
  (2003)

\bibitem{precisa}
Titolo, L., Feli{\'{u}}, M.A., Moscato, M., Mu{\~{n}}oz, C.A.: An abstract
  interpretation framework for the round-off error analysis of floating-point
  programs. In: Lecture Notes in Computer Science, pp. 516--537. Springer
  International Publishing (Dec 2017). \doi{10.1007/978-3-319-73721-8\_24},
  \url{https://doi.org/10.1007/978-3-319-73721-8\_24}

\end{thebibliography}

\end{document}